\documentclass[aos]{imsart}

\RequirePackage{amsthm,amsmath,amsfonts,amssymb}
\RequirePackage{natbib}
\RequirePackage[colorlinks,citecolor=blue,urlcolor=blue]{hyperref}
\RequirePackage{graphicx}
\usepackage{enumerate}
\usepackage{algorithm, algpseudocode}
\usepackage{url} 
\usepackage{caption}
\usepackage{subcaption}

\startlocaldefs

\newcommand{\piij}{p_{ij}}
\newcommand{\by}{\boldsymbol{y}}
\newcommand{\blambda}{\boldsymbol{\lambda}}

\newcommand{\boldzero}{\boldsymbol{0}}
\newcommand{\boldlambda}{\boldsymbol{\lambda}}
\newcommand{\bz}{\boldsymbol{z}}
\newcommand{\bzeta}{\boldsymbol{\zeta}}
\newcommand{\btheta}{\boldsymbol{\theta}}

\theoremstyle{plain}

\theoremstyle{remark}

\endlocaldefs

\begin{document}

\begin{frontmatter}
\title{Comparative Judgement Modeling to Map Forced Marriage at Local Levels}
\runtitle{Comparative Judgement Modeling to Map Forced Marriage at Local Levels}

\begin{aug}
\author[A]{\fnms{Rowland}~\snm{Seymour}\ead[label=e1]{r.g.seymour@bham.ac.uk}},
\author[B]{\fnms{Albert}~\snm{Nyarko-Agyei}}
\author[B]{\fnms{Helen}~\snm{McCabe}}
\author[C]{\fnms{Katie}~\snm{Severn}}
\author[C]{\fnms{David}~\snm{Sirl}}
\author[C]{\fnms{Theodore}~\snm{Kypraios}}
\and
\author[D]{\fnms{Adam}~\snm{Taylor}}
\address[A]{School of Mathematics, University of Birmingham\printead[presep={,\ }]{e1}}
\address[B]{Rights Lab, University of Nottingham\printead[presep={,\ }]{}}
\address[C]{School of Mathematical Sciences, University of Nottingham\printead[presep={,\ }]{}}
\address[D]{Digital Research Services, University of Nottingham\printead[presep={\ }]{}}
\end{aug}

\begin{abstract}
Forcing someone into marriage against their will is a violation of their human rights. In 2021, the county of Nottinghamshire, UK, launched a strategy to tackle forced marriage and violence against women and girls. We set out to map the risk of forced marriage across the county to support the strategy and enable the development of local interventions. However, there was no centralised database for forced marriage in the county and carrying out a survey using standard survey methods was unlikely to produce robust results due to the hidden nature of this crime. Comparative judgement provides a survey design that can map the risk of forced marriage through pair-wise comparisons. Current comparative judgement models require studies to have a large number of participants, so we developed a more flexible spatial modelling structure and a mechanism to schedule comparisons more effectively. The proposed modelling structure reduced the data collection burden and made a comparative judgement study feasible with a small number of participants. Underpinning this structure is a latent variable representation that improves on the scalability of inferential procedures of previous comparative judgement models. We used these methods to map the risk of forced marriage across Nottinghamshire, thereby supporting the county's strategy for tackling violence against women and girls.
\end{abstract}



\end{frontmatter}

\section{Introduction}


Forced marriage is a crime where one or both parties do not consent to the marriage and where pressure or abuse is used to force them into the marriage. It is a form of domestic abuse and a serious abuse of human rights, according to the UK Government, and is illegal under the Anti-social Behaviour, Crime and Policing Act (2014) and Marriage and Civil Partnerships (Minimum Age) Act (2022). Local level data on forced marriage allows the development of targeted safeguarding interventions and policies to protect victims of forced marriage; however, there is no publicly available data at such a level. Indeed, the UK Parliament’s Home Affairs Committee's inquiry into honour-based violence and forced marriage found the lack of data makes it difficult to ``formulate policy responses" to tackle honour-based abuse and forced marriage \citep{Home08}. Previous qualitative studies have shown local that local level policy on forced marriage in the UK is patchy, with procedures varying across local authorities \citep{Chantler2021}. The UK Government's Forced Marriage Unit is a specialised unit devoted to supporting victims and produces some statistics, but only provides spatial statistics at regional, not local, level.

We were motivated by the UK county of Nottinghamshire's strategy on tackling violence against women and girls \citep{Notts_VAWG}, which identified forced marriage as a key crime. One of the aims of the strategy is to work ``within the strategic assessments of local partnerships'' to tackle violence against women and girls. Our aims were to estimate the risk of forced marriage in each of the county's 76 wards and provide this information in a way that can inform such local partnerships. 

There are a wide variety of statistical models for estimating the location of hidden populations and analysing secondary or existing data using these models can be low-cost and straightforward. Examples of such data include casework data from social care or official statistics. However, unbiased data about forced marriage in Nottinghamshire does not exist and what data does exist, was inaccessible due to concerns about victim safety. In Nottinghamshire, there is no centralised database of forced marriage cases in the county. Victims are supported by a range of independent services (e.g. law enforcement, social services, voluntary agencies) and no one service supports all kinds of victims. Obtaining data-sharing agreements with all these services would be complex and time consuming, and services may be unwilling to share identifying information about victims.  We therefore decided to collect our own data through a bespoke survey. Surveys to locate victims of human rights abuses require careful formulation so that accurate and reliable information can be gathered without compromising victim anonymity. 

Comparative judgement models estimate the relative quality of objects in a set through pairwise comparisons. The survey's participants are shown pairs of objects from the set and asked which has the higher quality. People are, in general, more consistent at rating one object of a pair than rating an object on its own.  \citet{Jones2019} give the example of study participants finding it easier determine which of a pair of weights is heavier than determining the exact weight of an individual weight. In our motivating example, the \textit{objects} were \textit{wards}, the \textit{quality} was the \textit{risk} of forced marriage, and the \textit{participants}, referred to as \textit{judges}, were people who support victims of forced marriage and have knowledge of the victims' locations. This design allowed participants to share information without sharing details of victims or the exact number of victims they supported. We were able to use the data to map the relative risk of forced marriage without estimating prevalence directly. However, we needed to overcome two limitations to carrying out comparative judgement studies for measuring human rights abuses at local level.

Some existing comparative judgement models make use of spatial information in a limited way, assuming that neighbouring wards have similar qualities \citep[see, for example,][]{BSBT}. These models smooth the ward relative risk levels, providing a fine spatial structure about the risk of forced marriage at ward level. Ideally, the  local authorities would use a fine spatial structure to inform a ward-based approach to designing interventions. However, local authorities have limited financial resources and therefore need to take an approach that has lower cost than the tailoring interventions to each ward, but at the same still targets wards with similar levels of risk of forced marriage. For that reason, the county's aim is to design interventions based on ``local partnerships''.  We therefore developed a model that clusters ward by both geolocation and risk and apply this to inform the formation of such ``local partnerships''. This helped local authorities in Nottinghamshire to develop targeted interventions that have lower cost than interventions for each ward. That our clustering was done jointly on the basis of geography and relative risk allowed us to meet our and our partners' aims. Furthermore, when our model was fitted to data,the uncertainty in levels of relative risk was taken into account; something that most of off-the-shelf methods do not (e.g. a ward that has medium risk surrounded by high risk wards may be erroneously clustered with the high risk wards). 

A second limitation of existing comparative judgement methods is the need for a large number of study judges. The pool of potential judges in Nottinghamshire who knew about forced marriage numbered in the tens, whereas previous comparative judgement surveys used hundreds of judges (224 in \citet{BSBT} and 1,056 in \citet{Marshall20}). Deciding which objects should be compared has been considered before, but evaluation of survey designs has been limited due to high computational costs and intractability. \cite{Grass08} were able to analytically derive an experimental design but only for studies with at most three objects. Methods have been developed for studies where comparisons can be made in rounds, e.g. sporting contests, \citep{Glick05, Cattelan2012}, where after each round of comparisons, the next round is designed based on the information learnt so far. Our motivating example lacked a round structure, thus these methods are unsuitable. Adaptive comparative judgement has been used in education, which uses real time inference to estimate object qualities after each comparison \citep{Pollitt2012, Gray2024}. Objects with similar estimated qualities are more likely to be featured in a comparison, maximising the information learnt. We chose not to pursue this for two reasons. First, real-time inference \citep{Pollitt2012} is challenging and computationally expensive. Second, this would have caused an imbalance in the effort required by the judges, with the first judges being shown pairs at random and later judges being shown more difficult comparisons that required a higher cognitive load to complete. In this initial development of our methods, we prioritised ensuring that what we asked of volunteer judges is as similar as possible. \cite{Guo18} developed an experimental design but, due its high computational cost to be implemented, an approximate method for inference has been used. Therefore, we developed a scheduling mechanism for deciding which objects should feature in comparisons according to their geolocation and a scalable inference algorithm to obtain the efficiency of our tool. 



\subsection{Forced Marriage in Nottinghamshire}
\begin{figure}
    \includegraphics[width = 0.3\textwidth]{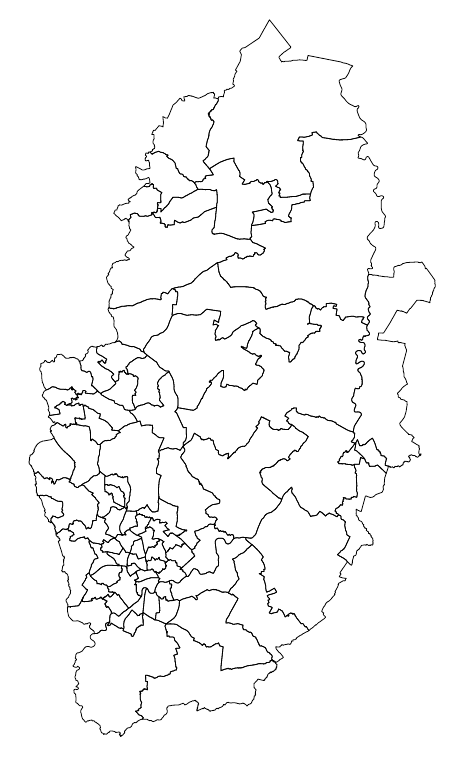}\qquad\qquad
    \includegraphics[width = 0.5\textwidth]{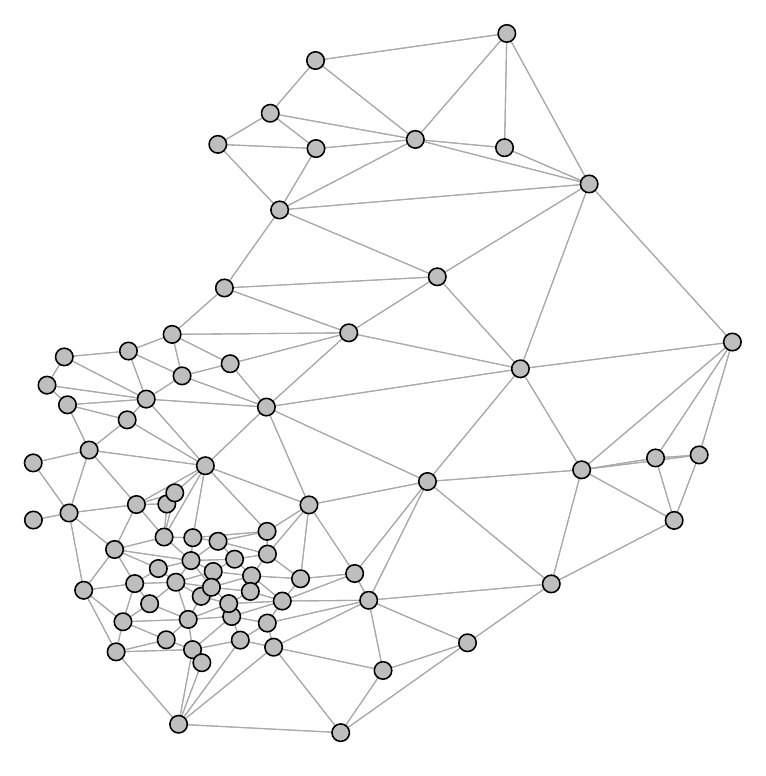}
    \caption{Left: A map of the upper tier local authority wards in Nottinghamshire. Right: The wards represented as a network, nodes represent wards and edges are placed between adjacent wards.}
    \label{fig: map of nottinghamshire}
\end{figure}

To map risk of forced marriage at local levels in Nottinghamshire, we carried out a comparative judgement study in the county. Nottinghamshire is a county in central England, with an approximate population of 750,000. Aside from the city of Nottingham in the south and the Mansfield conurbation in the west, the county is largely rural. The county consists of 76 local authority wards (20 in Nottingham City Council and 56 in Nottinghamshire County Council) -- a map of the wards used is shown in Figure \ref{fig: map of nottinghamshire}. 

We identified and recruited 12 judges who had county level expertise of forced marriage in Nottinghamshire. Judges were recruited via the Nottinghamshire Modern Slavery Partnership, which is a group of safeguarding professionals from law enforcement agencies, social services, education or the voluntary sector. Judges in the Partnership specialise in supporting different kinds of victims, for example children, adults or those with cognitive impairments. Judges typically work on a county level, meaning they can make comparisons about all wards in the county. Judges took part remotely via a web interface we designed; they were shown pairs of wards and asked to choose the ward with the lower rate per capita of forced marriage. Judges were able to say if they were not familiar with a ward, in which case the comparison was skipped and the ward was not featured in future comparisons. Judges were also able skip a comparison without a reason. We carried out a stakeholder mapping exercise to identify judges and emailed all judges with an explanation of the project and a link to the interface. We received ethical approval from the University of Nottingham School of Politics and International Relations ethics committee. 

\begin{figure}
    \centering
    \includegraphics[width = 0.8\textwidth]{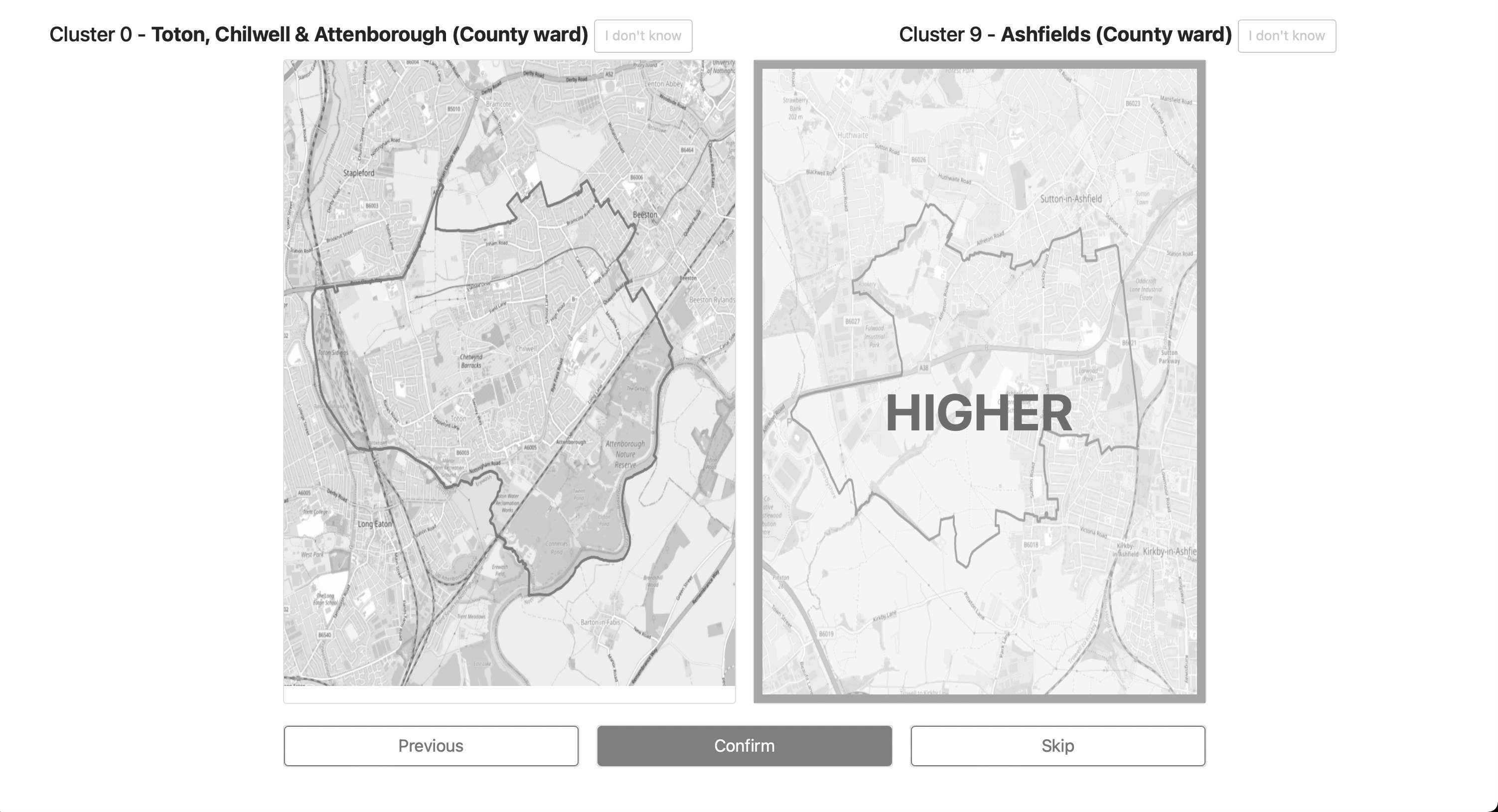}
    \caption{A screenshot of the web interface for the comparative judgement study. Here the judge was asked to compare Toton, Chilwell \& Attenborough ward with Ashfields ward. The judge chose Ashfields to have the higher rate per capita.}
    \label{fig: software}
\end{figure}

A web interface was developed for this study in Python version 3.9 and makes use of following external python packages; flask (v2.0.2), numpy (v1.22.0), pandas (v1.3.5), geopandas (v0.10.2) and SQLAlchemy (v1.4.29). Additionally, SQLite was used as the backend database during the study to collect and store comparative judgements. Images displayed to users during the assessment were sourced from OpenStreetMap data. It was based on the web interface used in \cite{BSBT}. An example of the web interface that judges see when participating in the study is shown in Figure \ref{fig: software}. 

\section{Spatial Bradley--Terry Models}

\subsection{The Standard Bradley--Terry Model}
One of the most widely used models for modelling comparative judgement data is the Bradley--Terry model \citep{Brad52} and is defined as follows. Consider a set of wards labelled $1, \ldots, N$. When comparing wards $i$ and $j$, the logit of the probability that ward $i$ is judged to have won the comparison is given by
\begin{equation}
    \hbox{logit}(\piij) = \lambda_i - \lambda_j \iff \piij = \frac{\exp(\lambda_i)}{\exp(\lambda_i) + \exp(\lambda_j)} \qquad (i \neq j, 1\leq i, j \leq N). \label{eq: logit difference}
\end{equation}

Typically, the parameters in the Bradley--Terry model, $\lambda_i$, $i=1,\ldots, N$ are interpreted as the ``quality'' or the ``ability'' of the objects of consideration. Hence, the parameter values are interpreted in a positive manner; i.e. the the higher the quality or ability, the better.  In the instructions that were provided in our study, the judges were given an example where if in ward $i$ there were two forced marriages for every thousand people, and in ward $j$ there were four for every thousand people, they should choose ward $i$. The ward with the lower rate (i.e. ward $i$ in this example) was coded as having won the comparison and perceived as a ward of ``higher quality''. In that sense, although the parameters encode some information about the rate of of forced marriage per capita in each ward, they are not, strictly speaking, {\em rate} parameters. Therefore, in our setting, $\lambda_i$ can be interpreted as the relative risk of forced marriage in ward $i$. 

 We recorded the outcome of each comparison between ward $i$ and $j$ as an 1 if ward $i$ was judged to have won the comparison and a 0 otherwise. We set $y_{ij}$ to be the sum of the outcomes for this pair of wards. To write the likelihood function for this model, let $n_{ij}$ be the number of times wards $i$ and $j$ are compared. Under the assumption that comparisons between different pairs of wards and also comparisons for a given pair are independent, the likelihood of the observed data is
\begin{equation}
  f(\by\mid\blambda) = \prod_{i=1}^N\prod_{j=1}^{i-1} \begin{pmatrix} n_{ij} \\ y_{ij}
\end{pmatrix} p_{ij}^{y_{ij}} (1-p_{ij})^{n_{ij} - y_{ij}},  \label{eq: BT likelihood}
\end{equation}
where $\boldsymbol{\lambda} = \{\lambda_1, \ldots, \lambda_N\}$ is the set of parameters describing the rates for forced marriage. 

Interpretation of the model parameters is then aligned with other comparative judgement studies where large positive values of the parameters $\lambda$ correspond to high quality wards (i.e. wards with low risk of forced marriage) and large negative values of $\lambda$ corresponding to low quality wards (i.e. wards with a high risk of forced marriage) There are a wide variety of methods for estimating the model parameters, with both classical and Bayesian methods being available \citep{Turner2012, Cattelan2012}. As our application has a spatial element, we briefly describe the Bayesian Spatial Bradley--Terry model (BSBT) \citep{BSBT}. 

\subsection{The Bayesian Spatial Bradley--Terry Model} \label{sec: BSBT}
The BSBT model allows for the incorporation of prior assumptions about spatial structure of the risk parameters $\blambda$. A zero-mean multivariate normal distribution is placed on the risk parameters
\begin{equation}
  \blambda \sim \rm MVN(\boldzero, \Sigma).
  \label{eq: prior}
\end{equation}

The covariance matrix $\Sigma$ encodes prior assumptions about the spatial structure of the risk parameters. To construct the covariance matrix $\Sigma$ in our study, we created a network from the wards in the study, treating the wards as nodes and placing edges between adjacent wards. See Figure \ref{fig: map of nottinghamshire} for the network constructed from the wards in Nottinghamshire. We let $\Lambda = e^A$, where $A$ is the adjacency matrix of the network, and $D$ be the matrix containing the diagonal elements of $\Lambda$. The prior distribution covariance matrix was given by
$$
\Sigma = \alpha^2 D^{-\frac{1}{2}}\Lambda D^{-\frac{1}{2}},
$$
where $\alpha^2$ is a signal variance parameter. Using the matrix exponential assigns high correlation to pairs of wards that are well connected and low correlation to pairs that are only connected via long paths. The normalisation by $D$ ensures the diagonal elements of $\Sigma$ are $\alpha^2$ and the off-diagonal entries are proportional to the communicability of each pair of subwards in the network \citep{Estrada2010}. We placed a conjugate inverse-Gamma prior distribution on the signal variance parameter $\alpha^2$ with shape $\chi$ and scale $\omega$. 

The posterior distribution for this model is given by
$$
\pi(\blambda, \alpha^2 \mid \by) \propto f(\by \mid \blambda)\,\pi(\blambda \mid \alpha^2)\,\pi(\alpha^2). 
$$

\subsection{Spatial Clustering Bradley--Terry Model} \label{sec: clustering}
To address the aim of Nottinghamshire's strategy to produce results that can inform local partnerships, we developed a model that clusters the wards by geolocation and risk. For such problems, it is possible to specify the number of clusters \textit{a priori}, but it is difficult to justify such a choice. We utilised a Bayesian nonparametric method known as a distance dependent Chinese restaurant process (ddCRP) \citep{Blei11}, and we developed a model where the number of clusters was not specified in advance, but instead learned as part of the statistical inference. 

In the spatial clustering BT model, each ward is associated with one other ward (which could be itself) and the clusters are defined implicitly through these associations. The risk parameters $\blambda$ are independently and identically distributed according to an underlying distribution for each cluster. We assume the distribution of the qualities of wards in cluster $k$ is a $N(m_k, \sigma^2_k)$ distribution, and that there is a base distribution $G_0$, according  to which the cluster mean and variance parameters are distributed. We take this to be the normal inverse-gamma distribution. We let the assignment of ward $i$ be denoted by $\theta_i$, and the prior distribution on $\theta_i$ is  
$$
p(\theta_i = j\mid \beta) \propto \begin{cases}
h(i, j) &\textrm{if } i\neq j \\
\beta &\textrm{if } i= j. \\
\end{cases}
$$
The concentration parameter $\beta$ controls how likely each ward is to associate with itself, with large values of $\beta$ giving many smaller clusters and small values of $\beta$ giving fewer, larger clusters. The function $h$ determines the probability pairs of wards are associated with each other and is a modelling choice. In our application, we used a distance based metric based on the graph distance in the network of the wards. The posterior distribution for this model is \begin{align*}
\pi(\boldlambda, \btheta, \boldsymbol{m}, \boldsymbol{\sigma}^2\mid \by) \propto f(\by \mid \boldlambda)\, f(\boldlambda \mid \boldsymbol{m}, \boldsymbol{\sigma}^2, \btheta)\,\pi(\btheta) \,\pi(\boldsymbol{m} \mid \boldsymbol{\sigma}^2)\, \pi(\boldsymbol{\sigma}^2),
\end{align*}
where $\btheta, \boldsymbol{\mu}$ and $\boldsymbol{\sigma}^2$ are sets containing the ward assignments, cluster mean and variance parameters respectively. 

\section{Efficient Data Collection for Spatial Bradley--Terry Models} \label{sec: experimental design}
The second limitation we identified was the limited number of judges. When collecting comparative judgement data about specific abuses or at fine grain levels, there may be a limited number of judges who have sufficient expertise to take part in the study. Therefore, it is particularly important to ask judges to make comparisons that are likely to maximise the information gained in the study. In our motivating example, we were unable to control the number of judges or number of comparisons. Instead, we controlled the comparisons the judges are asked to make, showing them comparisons that elicited the most information. 

In a comparative judgement study, the schedule $\boldsymbol{s}$ is the set of $M$ upcoming comparisons. We focus on schedules where upcoming comparisons are drawn from a distribution, although it is possible to use schedules which are deterministic or manually created. Previous studies have not considered spatial structure when scheduling comparisons. For example, in \cite{BSBT} comparisons were scheduled uniformly at random from all possible pairs of wards. 

In our motivating study, asking judges to compare two highly connected wards may not have elicited the most informative response, as we assumed the rates are highly correlated. We pursued a scheduling method that utilises the prior covariance structure, with pairs of wards that are not highly connected being prioritised over pairs of wards that are. We derived a probability distribution over all possible comparisons to represent this mechanism. When scheduling comparisons for a study, the pairs of wards are to be compared are drawn from this distribution. 

\subsection{Scheduling Comparisons for a Spatial Comparative Judgement Study}
To construct our scheduling distribution, we adapted a method from principal component analysis to weight the importance of pairs of wards by the amount of prior variance they explain.  We denote the distribution by $\mathcal{S}$ and assume $s_m \overset{iid}{\sim} \mathcal{S},\, m = {1, \ldots, M}$, is the labels of the $m^{th}$ pair of wards to be compared. 

As comparisons provide information about one ward relative to another, instead of absolute information, we construct the distribution of  difference in the risk parameters. This distribution can be obtained through an affine transformation of the prior distribution $\blambda \sim N(\boldsymbol{\mu}, \Sigma)$ and is given by $\blambda_{\rm diff} \sim N(\boldsymbol{\nu}, \Delta)$. The vector $\blambda_{\rm diff} = \{\lambda_1 - \lambda_2, \lambda_1 - \lambda_3, \ldots, \lambda_{N-1} - \lambda_{N}\}$ contains the difference in qualities of each pair of wards, $\boldsymbol{\nu}$ is the corresponding vector of differences in the mean parameters and $\Delta$ is the matrix containing the covariances between each pair of pairs of relative differences. The element of $\Delta$ corresponding to the covariance between the pair $(i, j)$ and $(k, l)$ is given by
$$
\hbox{Cov}(\lambda_i - \lambda_j, \lambda_k - \lambda_l) = \hbox{Cov}(\lambda_i , \lambda_k) - \hbox{Cov}(\lambda_i , \lambda_l) - \hbox{Cov}(\lambda_j , \lambda_k) + \hbox{Cov}(\lambda_j , \lambda_l).
$$
The spectral decomposition of the covariance matrix $\Delta$ is $\Delta=U\Psi U^T$, where $\Psi$ is a diagonal matrix of eigenvalues of $\Delta$ and columns of $U$ are the corresponding eigenvectors.  We order the eigenpairs $\{\psi^{(c)}, \boldsymbol{u}^{(c)}\}$ such that $\psi^{(c)} \geq \psi^{(c+1)}$ (recall that $\psi^{(c)} \geq 0$ since $\Delta$ is positive semidefinite). The eigenvectors create an orthogonal basis for $\mathbb{R}^{N(N-1)/2}$ and each vector is known as a principal component \citep{Mardia79}. 

The $c^{th}$ principal component is the vector that explains the maximum variance of the prior distribution of $\blambda_{\rm diff}$ whilst still being orthogonal to the $c-1$ vectors before it. The proportion of variance explained by the $c^{th}$ principal component is $\frac{\psi^{(c)}}{\sum_d \psi^{(d)}}$ and the $i^{th}$ variable in the $c^{th}$ principal component can be thought of as explaining $\frac{\left(\mathbf{u}^{(c)}_i\right)^2}{\sum_k\left(\mathbf{u}^{(c)}_k\right)^2}$ proportion of this proportion of variability.

We then define the probability pair $\{i, j\}$ is shown to a judge is
\begin{align*}
    q_{ij} = \frac{\sum_c\left(\mathbf{u}^{(c)}_{r}\right)^2\psi^{(c)}}{\sum_c\psi^{(c)}}, 
\end{align*}
with $r = \frac{N(N-1)}{2} - \frac{(N-i +1)(N-i)}{2} + j - i$ being the linear index of $\blambda_{\rm diff}$ corresponding to the pair $(i, j)$. This is equivalent to the sum of the loadings squared for a ward's risk parameter over the total variance explained. The set $\{q_{1,2}, \ldots, q_{N,(N-1)}\}$ describes the probability distribution $\mathcal{S}$ over the set of pairs of wards which places higher mass on pairs which have higher prior variance. 

\section{Scalable Bayesian Inference for Bradley--Terry Models}
We required a scalable inference algorithm to evaluate the utility of the scheduling mechanism. We now describe our method to perform posterior computation, a latent variable algorithm. Our latent variable algorithm is considerably more efficient than the currently available algorithms \citep{BSBT}, providing a better mixing Markov chain and a faster time to convergence. 

\subsection{P\'olya-Gamma Latent Variable Representation} \label{sec: PG}
We now present an alternative latent variable formulation of the BT model that allows incorporation of a prior covariance structure for the risk parameters $\blambda$ and leads to a very efficient Gibbs sampler; this formulation is based on \citet{Caron2012}. Consider the likelihood contribution from all comparisons of the  pair of wards $(i, j)$. If this pair is compared $n_{ij}$ times,  out of which  ward $i$ was judged to be superior to ward $j$ $y_{ij}$ times, then the likelihood contribution is
$$
f_{ij}(\blambda) = \frac{\exp\left(\boldsymbol{x}_{ij}^T \,\boldsymbol{\lambda}\right)^{y_{ij}}}{\left(1 + \exp\left(\boldsymbol{x}_{ij}^T \,\boldsymbol{\lambda}\right)\right)^{n_{ij}}}, \quad i,j= 1,\ldots, N,
$$
where $\boldsymbol{x}_{ij}$ is $N \times 1$ vector with all elements being zero apart from the $i^{\mbox{\tiny{th}}}$ and $j^{\mbox{\tiny{th}}}$ elements which take values $1$ and $-1$ respectively. Introducing a latent variable $z_{ij}$ which follows a P\'olya-Gamma distribution and then following \citet{Polson2013} we can write
\begin{equation*}
    f_{ij}(\blambda) \propto \exp\left(k_{ij}\, \boldsymbol{x}_{ij}^T \,\boldsymbol{\lambda}\right) \int_{0}^{\infty}\exp\left(- z_{ij} \left(\boldsymbol{x}_{ij}^T \,\boldsymbol{\lambda}\right)^2/2\right) f(z_{ij}|n_{ij}, 0) \mbox{ d}z_{ij}, 
\end{equation*}
where $k_{ij} = y_{ij} - n_{ij}/2$ and $f(z_{ij}|n_{ij}, 0)$ is the density of a P\'olya-Gamma random variable with parameters $n_{ij}$ and 0. We can now augment the observed data $\boldsymbol{y}$ from all $M$ pairs of wards with the corresponding set $\boldsymbol{z} = (z_{ij})$ of latent variables for all pairs of wards and write the observed-data likelihood as follows:
\begin{align}
f(\by|\blambda) & \propto  \int_{\boldsymbol{z}} f(\by, \bz|\blambda) \mbox{ d}\bz = \prod_{i=1}^{N}\prod_{j<i} f_{ij}(\blambda) \nonumber \\
& =  \prod_{i=1}^{N}\prod_{j<i} \Bigg( \exp\left(k_{ij}\, \boldsymbol{x}_{ij}^T \,\boldsymbol{\lambda}\right) \int_{0}^{\infty}\exp\left(- z_{ij} \left(\boldsymbol{x}_{ij}^T \,\boldsymbol{\lambda}\right)^2/2\right) f(z_{ij}|n_{ij}, 0) \mbox{ d}z_{ij}\Bigg).    
\end{align}
As in the Exponential latent variable representation in \citet{Caron2012}, the introduction of $\bz$ is instrumental in constructing a scalable and efficient MCMC algorithm. The conditional posterior distributions $\pi(\bz | \by, \blambda)$ and $\pi(\blambda | \bz, \by)$ are available in closed form, leading to a straightforward implementation of a Gibbs sampler. Letting $X$ be the BT design matrix constructed from $\boldsymbol{x}_{ij}$, the conditional distribution of $\blambda|\by, \bz$ is derived by
\begin{align*}
\pi(\blambda|\bz, \by) & \propto \pi(\blambda) \prod_{i=1}^{N}\prod_{j<i} f_{ij}(\blambda) \nonumber \\
& =  \pi(\blambda)\prod_{i=1}^{N}\prod_{j<i} \exp\left(k_{ij}\, \boldsymbol{x}_{ij}^T\,\boldsymbol{\lambda} - z_{ij} \left(\boldsymbol{x}_{ij}^T \,\boldsymbol{\lambda}\right)^2/2 \right) \nonumber \\
& = \pi(\blambda) \prod_{i=1}^{N}\prod_{j<i}\exp\left(\frac{z_{ij}}{2} \left(\boldsymbol{x}_{ij}^T\, \blambda - \frac{k_{ij}}{z_{ij}} \right)^2 \right) \nonumber \\ 
& = \pi(\blambda) \exp\left(-\frac{1}{2}\left(\bzeta - X\blambda \right)^T Z (\bzeta - X \blambda) \right) \nonumber \\
& = \pi(\blambda) \exp \Bigg( - \frac{1}{2} \left(\blambda - X^{-1} \bzeta\right)^{T} \left( X^T Z X \right)  \left(\blambda- X^{-1}\bzeta \right)\Bigg), \nonumber 
\end{align*}
where $Z = \mbox{diag}(z_{ij})$ and $\bzeta = \left( k_{12}/z_{12}\,, k_{13}/z_{13}, \ldots, k_{N-1\, N}/z_{N-1 \, N} \right)$. Assigning a multivariate Normal prior distribution on $\blambda$  that allows for a priori dependence among the risk parameters $\blambda$, i.e. $\blambda \sim N(\bar{\boldsymbol{\mu}}, {\Sigma})$, it can be shown that
\begin{equation}
\blambda|\by, \bz \sim N(\boldsymbol{\mu}, B), \label{eq:conditional_lambda}
\end{equation}
where $B = \left(X^{T} Z X + \Sigma^{-1} \right)^{-1}$ and $\boldsymbol{\mu} = B \left(X^T \left(\by -\boldsymbol{n}/2)\right) + \Sigma^{-1}\bar{\boldsymbol{\mu}}\right)$. By the definition of a P\'olya-Gamma random variable and its expectation \citet{Polson2013} show that
\begin{equation}
z_{ij}|y_{ij}, \lambda_{ij} \sim PG(n_{ij},\lambda_{i} - \lambda_{j}). \label{eq:z_given_lambda}
\end{equation}
Sampling from (\ref{eq:z_given_lambda}) can be efficiently done using an accept/reject algorithm based on the partial, alternating-series method of \citet{Devroye1986}. In conjunction with exploiting the sparse structure of the matrix $X$, we have a Gibbs sampler in Algorithm \ref{alg: PG} that allows sampling of the joint posterior distribution of $\blambda$ and $\bz$ and which is scalable and efficient to implement.

\begin{algorithm}[h]
\caption{The P\'olya-Gamma latent variable algorithm }
	\begin{algorithmic}
		\State Initialise the chain with values $\lambda_1, \ldots, \lambda_m$ and $z_{1, 2} \ldots z_{N-1, N}$
		\Repeat
        \For{$1 \leq i < j \leq N, n_{ij} > 0$}
            \State  Draw $z_{ij} \sim PG(n_{ij}, \lambda_i - \lambda_j)$
        \EndFor
        \State Draw $\blambda \sim \textrm{MVN}(\boldsymbol{\mu}, B)$
        \State Draw values for any model hyperparameters
        \Until{Markov chain has converged}
	\end{algorithmic}
	\label{alg: PG}
\end{algorithm}

An R software package to implement the MCMC algorithm can be found at \url{https://github.com/rowlandseymour/speedyBBT}.

\subsection{Spatial Clustering Bradley--Terry Model}
\begin{figure}
     \centering
     \begin{subfigure}[b]{0.3\textwidth}
         \centering
         \includegraphics[width=\textwidth]{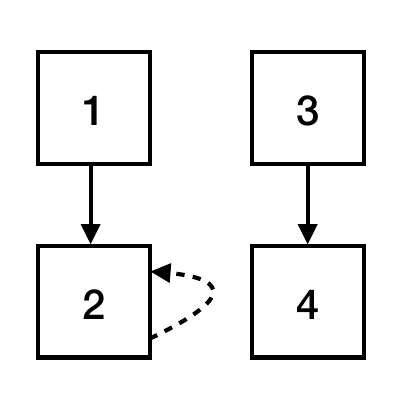}
         \caption{Ward 2 is assigned to itself creating two clusters.}
         \label{fig: ddcrp current}
     \end{subfigure}
     \hfill
     \begin{subfigure}[b]{0.3\textwidth}
         \centering
         \includegraphics[width=\textwidth]{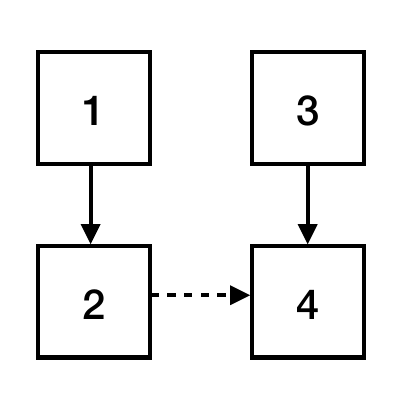}
         \caption{Ward 2 is assigned to ward 4, creating one  cluster.}
         \label{fig: ddcrp join}
     \end{subfigure}
     \hfill
     \begin{subfigure}[b]{0.3\textwidth}
         \centering
         \includegraphics[width=\textwidth]{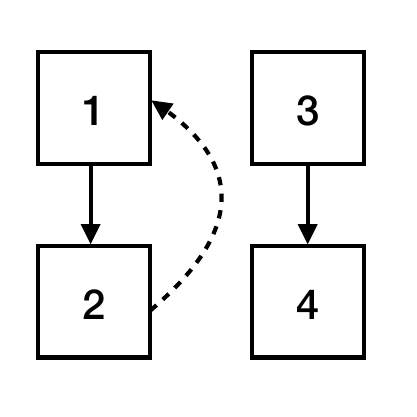}
         \caption{Ward 2 is assigned to ward 1, creating two  clusters.}
         \label{fig: ddcrp dont join}
     \end{subfigure}
        \caption{An example of three configurations when assigning ward 2 to another ward. The dashed arrow corresponds to the assignment of ward 2. }
        \label{fig: ddcrp example}
\end{figure}

The P\'olya-Gamma latent variable representation can be used with the spatial clustering model and we now adapt Algorithm \ref{alg: PG} to construct an MCMC algorithm for this model. Recall that in the clustering model each ward is assigned to another ward $\theta_i$ and $\boldsymbol{\theta} = (\theta_i)$ define the clustering.  When updating $\theta_i$ there are three possibilities we need to consider: that $\theta_i = i$; that $\theta_i = j \neq i$ and this assignment results in two clusters merging; that that $\theta_i = j \neq i$ and this connection does not result in two clusters merging.  Example configurations are given in Figure \ref{fig: ddcrp example}. Denote the two clusters that $\theta_i$ could link by cluster $k$ and cluster $l$, where $\boldsymbol{\lambda}_k$ and $\boldsymbol{\lambda}_l$ are the sets of risk parameters of ward in cluster $k$ and $l$ respectively. The full conditional distribution for $\theta_i$ is 
$$
\pi(\theta_i \mid \btheta_{-i}, \boldsymbol{\lambda}^*) \propto
\begin{cases}
\beta & \hbox{if } \theta_i = i,  \\
h(i, j) \frac{f(\boldsymbol{\lambda}_k , \boldsymbol{\lambda}_l | G_0)}{f(\boldsymbol{\lambda}_k | G_0) f(\boldsymbol{\lambda}_l | G_0)} & \hbox{if } \theta_i = j \hbox{ and clusters } k \hbox{ and } l \hbox{ would be joined},\\
h(i, j) & \hbox{if } \theta_i = j \hbox{ and a two clusters would not be joined,} \\
\end{cases}
$$
where $\btheta_{-i}$ is the set of assignments excluding the assignment for ward $i$, and $\boldlambda_k$ are the risk  parameters associated with wards in cluster $k$. The term $f(\boldsymbol{\lambda}_k | G_0)$ is the marginal likelihood evaluated at the the risk parameters in cluster $k$:
\begin{align*}
    f(\boldsymbol{\lambda}_k \mid G_0) &= \iint \prod_{i; \theta_i = k} \pi(\lambda_i \mid \mu, \sigma^2)\, dG_0(\mu, \sigma^2) \\
    &= \iint \left(\prod_{i; \theta_i = k} \pi(\lambda_i; \mu, \sigma^2) \right) \pi(\mu | \sigma^2) \pi(\sigma^2)\, d\mu\, d\sigma^2 \\
    & = \frac{\Gamma(\alpha_o + n_k/2)}{\Gamma(\alpha_0)}\frac{\beta_0^{\alpha_0}}{\bar{\beta}^{\alpha_0 + n_k/2}}\left(\frac{1}{1 + n_k}\right)^\frac{1}{2} (2\pi)^{-\frac{n_k}{2}},
 \end{align*}
where $\bar{\beta} = \beta_0 + \frac{1}{2}\sum_{i; \theta_i = k}(\lambda_i - \theta_k )^2 + \frac{n_k(\theta_k - \mu_0)^2}{2(1 + n_k)}$ and $\theta_k = \frac{1}{n_k}\sum_{i; \theta_i = k}\lambda_i$. The term $f(\boldsymbol{\lambda}_k, \boldsymbol{\lambda}_l | G_0)$ is the marginal likelihood evaluated at all the risk parameters in clusters $k$ and $l$.

We parameterise the cluster variance parameters through their inverses, the cluster precision parameters. The full conditional distribution for the precision of cluster $k$ is 
$$
\frac{1}{\sigma^2_k} \mid \boldsymbol{m}, \boldlambda, \btheta \sim \Gamma\left(\alpha_0 + \frac{n_k}{2}, \, \bar{\beta}\right). 
$$
The full conditional distribution for the mean parameter for cluster $k$, $\lambda_k^*$, is given by
$$
m_k | \sigma^2_k, \boldlambda, \btheta \sim N\left(\frac{\mu_0 + n_k \theta_k}{1 + n_k}, \frac{\sigma^2_k}{1 + n_k}\right). 
$$


\subsection{Identifiability}
Model (\ref{eq: logit difference}) is invariant to translations and hence an identifiability constraint is necessary when inferring the risk parameters $\blambda$. This does not change the value of the likelihood function as it is invariant to translations. We adapt the translation in \citet{Caron2012} and define 
$$
\Lambda = \frac{1}{N}\sum_{i=1}^N \lambda_i
$$
to be the mean of the risk parameters. Under the prior distribution (\ref{eq: prior}) the distribution of $\Lambda$ is 
$$
\Lambda \sim N\left(0, \frac{\boldsymbol{1}\Sigma\boldsymbol{1}^T}{N^2}\right),
$$
where $\boldsymbol{1} = (1, \ldots, 1)^T$ is a vector of ones. At each iteration of Algorithm  \ref{alg: PG} we draw a value for $\Lambda$ and translate the values of $\blambda$. We found this to greatly improve the mixing of our MCMC algorithm and  by doing so, we are also enable to do fair comparison  against the sampler of in \cite{Caron2012}.

\section{Simulation Studies} \label{sec: sim study}

\subsection{Scalability of the latent variable representation}
A framework for scalable inference is essential to be able evaluate the utility of our experimental design. To evidence the scalability of the P\'olya-Gamma latent variable representation in spatial studies, we simulated synthetic parameter values for the risk of forced marriage in Nottinghamshire and inferred them using (i) the P\'olya-Gamma latent variable representation, (ii) a single-site sampler with an independent proposal distribution based on the MLE and quasi-variance for each risk parameter, and (iii) a block update Metropolis-Hastings random walk algorithm \citep{BSBT}. We simulated 100 sets of risk values for the wards of Nottinghamshire by drawing from the prior distribution. For each set of parameter values, we simulated 2,000 comparisons using the scheduling mechanism described in Section 3, as this closely matches the number of comparisons we received in the study.

To compare the computational efficiency of the different samplers, we computed the effective sample size per second (ESS/s). Across all sets of comparisons and sets of risk parameters, the median ESS/s for the P\'olya-Gamma latent variable representation is 60.03 (min: 9.978, max: 97.6), compared to 1.175 (0.087, 44.7) for the sampler used in \cite{BSBT} and 2.94 (0.83, 5.04) for the single-site sampler. This represents a fifty-fold and twenty-fold increase in the efficiency on average compared to the sampler with the Metropolis-Hastings random walk and the single site samplers respectively. More information on this simulation study and another study, where the parameter values of the risk parameters are drawn independently, can be found in the Supplementary Material. 

\subsection{Accuracy of the clustering model}
We carried out a further simulation study to investigate the ability of the clustering model to recover cluster assignments and the values of the risk parameters. We simulated 50 sets of cluster assignments for the wards in Nottinghamshire using the concentration parameter value $\beta= 1\times10^{-8}$ and setting the distance metric to be $h(i, j) = \exp(A)_{ij}$, where $A$ is the adjacency matrix of the wards in the county. This particular value for the concentration parameter produced datasets which had between one and eight clusters which is reasonable for the local authorities in Nottinghamshire. For each set of clustering assignments, we simulated a set of risk parameters according to the clustering model and then simulated 2,000 comparison for each set of risk parameters, which is similar to the number of comparisons collected in the real data. Full details of the parameter values used can be found in the Supplementary Material. We fitted the clustering model to each of the 50 sets of comparisons and learned the cluster assignments and the risk parameters. We computed the {\em rand} and {\em adjusted rand} indices between the simulated and learned cluster assignments \citep{Rand1971}, to assess the accuracy of the model in learning the cluster assignments. These indices measure the similarity in the cluster assignments of the wards on a pairwise basis, with 1 corresponding to complete agreement. Across the 50 sets of comparisons, the median indices were 0.917 and 0.800 for the rand and adjusted rand index respectively, suggesting our model is able to accurately learn the clustering assignments. We also computed the median absolute error in the risk parameters for each set of comparisons. The median value of this index across all 50 data sets was 0.265, suggesting we can indeed learn the risk parameters using our proposed model. 

\subsection{Utility of the scheduling mechanism}
We assessed the utility of our spatial scheduling mechanism and compared it against two other scheduling mechanisms. The first scheduling mechanism was where pairs of wards are chosen uniformly at random from all possible pairs of wards. The second was a naive spatial scheduling mechanism, where the probability a pair of wards is chosen for comparison depends on how connected they are. To construct this mechanism, firstly, we constructed a vector $\boldsymbol{p}^*$, where each element describes the connectedness between a pair of wards, i.e. the matrix $\exp(A)_{ij}$ in vector form. Secondly, we normalised $\boldsymbol{p}^*$ to create a probability distribution. The distribution governing the scheduling mechanism is given by
$$
\boldsymbol{p}^*_{\textrm{norm}} \propto \boldsymbol{1} - \frac{\boldsymbol{p}^*}{\sum \boldsymbol{p}^*}. 
$$
We subtracted the normalised vector from one to assign low probability to pairs of wards that are highly connected and high probability to pairs of wards that are not.

We simulated 1,000 synthetic sets of risk parameters for the wards in Nottinghamshire. The values for the risk parameters were drawn from the distribution in equation (\ref{eq: prior}) with $\alpha = 3$. For each of the sets of risk parameters we simulated three sets of 500 comparisons, each corresponding to one of the three scheduling mechanisms. We chose 500 comparisons, as this represents 10 judges doing 50 comparisons each, which was our target for the data collection in out study.  We then fitted the model to the simulated comparisons using the P\'olya-Gamma latent variable representation and ran the algorithm for 500 iterations; we removed the first 50 iterations as a burn-in period. For each set of comparisons we recorded the utility of the scheduling mechanism. Running such a large number of simulations is only computationally feasible using the P\'olya-Gamma latent variable representation,

To assess the effectiveness of these different schedules, we defined two utility functions that measure different aspects of our aim. There exist a wide range of utility functions \citep[see, e.g.,][]{Ryan2015}, but due to the identifiability issues in the BT model, our choice of utility function is limited. Consider a schedule $\boldsymbol{s} \in \mathcal{S}$ given the comparisons $\boldsymbol{y}$. The first function we defined was the Bayesian A-posterior precision utility function \citep{Ryan2015}
$$
U_1(\boldsymbol{s} \mid \boldsymbol{y}) = \frac{1}{\textrm{tr}(\textrm{cov}(\boldsymbol{\lambda} \mid \boldsymbol{s}, \boldsymbol{y}))},
$$
where $\textrm{cov}(\boldsymbol{\lambda} \mid \boldsymbol{s}, \boldsymbol{y})$ is the posterior covariance matrix of the parameters $\blambda$. The utility function depends on the precision of the marginal posterior distributions of the risk parameters, which measures our ability to learn the risk of forced marriage in each ward. As we were also interested in rank ordering of each pair of wards is accurate, we constructed a second utility function that depends on the probability one ward is judged to have a higher risk than another. We defined the second utility function as 
$$
U_2(\boldsymbol{s} \mid \boldsymbol{y}) = \frac{2}{N(N-1)}\sum_{i=1}^N\sum_{j < i} \frac{1}{\hbox{Var}(\pi(p_{ij} \mid \boldsymbol{y}))},
$$
where $\pi(p_{ij} \mid \boldsymbol{y})$ is the posterior predictive distribution of the probability ward $i$ is judged to have a higher rate of forced marriage than ward $j$ given the comparisons $\boldsymbol{y}$. This utility function assesses the average precision in our ability to predict which of a pair of wards has a higher risk of forced marriage, and should therefore be ranked higher. 

Tables \ref{tab: design sim study U1} and \ref{tab: design sim study U2} summarise the utility across all 5,000 sets of comparisons. The scheduling mechanism based on principal components had a greater utility than either of the other two deigns, with the mean utility for this scheduling mechanism around 13\% higher (as measured by $U_1$) and 10\% higher (as measured by $U_2$) than for the two simpler scheduling mechanisms. This mechanism also allowsed for more information to be gained from fewer comparisons, and, on average, provided an increase in the utility of the scheduling mechanism. Despite taking the spatial structure into account, the naive spatial scheduling mechanism had the same utility as the uniform scheduling mechanism, evidencing the effectiveness of our new spatial scheduling mechanism at gaining information from judges. 

\begin{table}
\centering
    \caption{The mean, minimum, and maximum utility ($U_1$) for the three scheduling mechanisms across 5,000 simulated data sets.   \label{tab: design sim study U1}}
    \begin{tabular}{|c|c|c|c|} \hline
       Scheduling mechanism & Mean  & Minimum & Maximum \\ \hline
       Uniform & 0.199 & 0.106 & 0.256\\
       Naive spatial  & 0.197 &  0.108 & 0.257 \\
       Principal component  & 0.224 & 0.116 &  0.292 \\ \hline
    \end{tabular}
\end{table}

\begin{table}
\centering
    \caption{The mean, minimum, and maximum utility  ($U_2$) for the three  scheduling mechanisms across 5,000 simulated data sets.   \label{tab: design sim study U2}}
    \begin{tabular}{|c|c|c|c|} \hline
       Scheduling mechanism & Mean  & Minimum & Maximum \\ \hline
       Uniform &  864 & 316 &  40011\\
       Naive spatial  &  824 & 314 & 53935\\
       Principal component   & 952 &   337 & 43427 \\ \hline
    \end{tabular}
\end{table}

\section{Forced Marriage in Nottinghamshire} \label{sec: Notts}
There were a limited number of individuals with sufficient expertise in forced marriage in Nottinghamshire to act as judges in this study, and this was our motivation to develop a method to maximise the information gained from each comparison. We identified 29 organisations supporting victims of forced marriage in the county, this included large organisations such as local authorities and small organisations, such as charities run by volunteers. The organisations were identified through a stakeholder mapping exercise and through the Nottinghamshire modern slavery partnership (forced marriage has been included as a form of modern slavery by the International Labour Organisation since 2017).

Invites to take part in the study were emailed to individuals from all 29 organisations identified. We recruited 12 judges who made 1,848 comparisons. To ensure that no judge could be identified by their comparisons, no data was collected about the judges and each judge was assigned a random ID number when they enrolled in the study. Table \ref{tab: number comparisons} shows the number of comparisons each judge made.

\begin{table}

        \caption{Summary of the judges and comparisons they made in the study.\label{tab: number comparisons}}
    \begin{tabular}{|c|c|c|c|c|c|c|c|c|c|c|c|c|} \hline
    Judge &    1 &  2 &  3  & 4   & 5   & 6 & 7  & 8  & 9  & 10 & 1  & 12 \\ \hline
 \# comp. &  11  & 27 & 62  & 344 & 131 & 1 & 36 & 81 & 8  & 3 &  81 & 1063  \\ \hline
 Median time taken (s) &  7.5  & 8.5 & 5.5  & 4 & 8 & - & 7.5 & 6 & 11.5  & 3 &  4 & 4  \\ \hline
    \end{tabular}
\end{table}

\begin{figure}
    \centering
    \includegraphics[width = 0.5\textwidth]{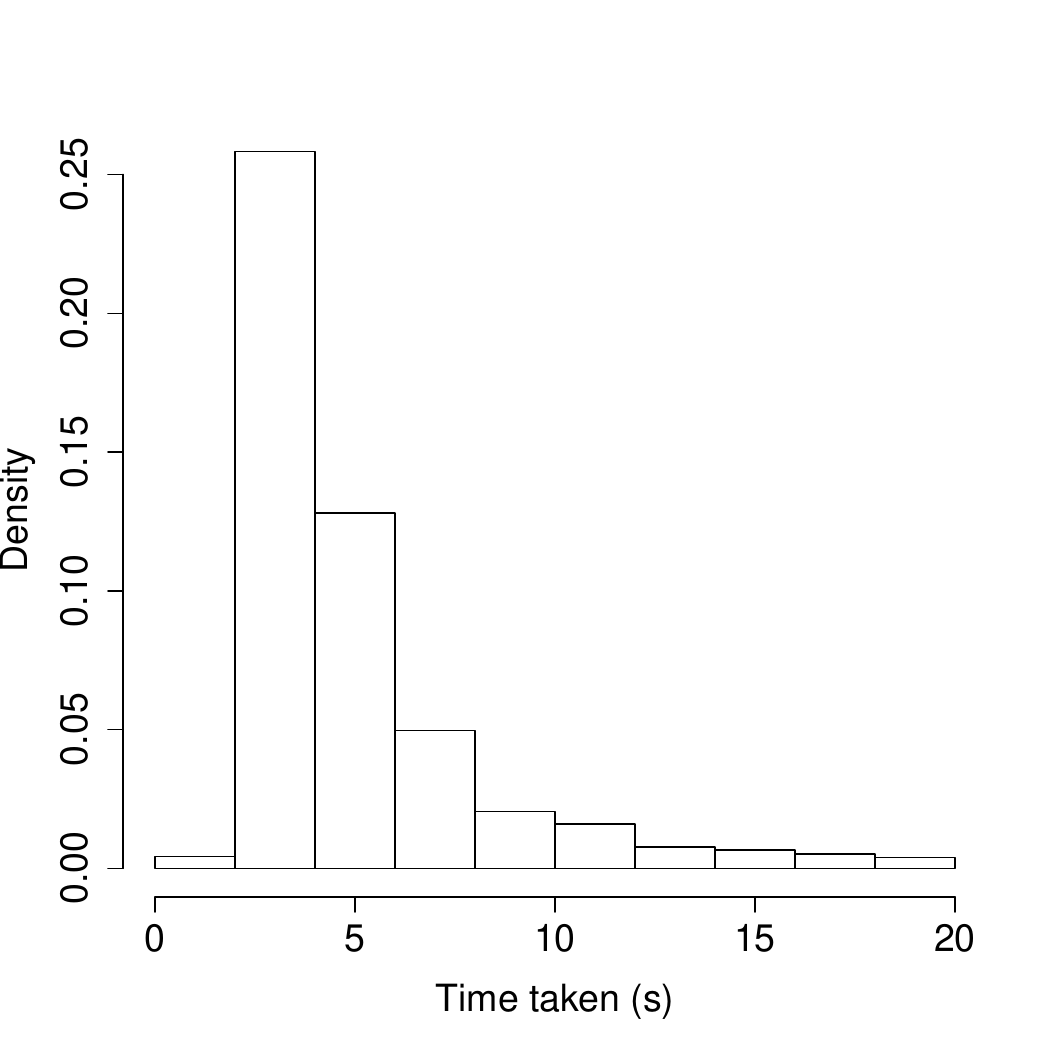}
    \caption{The distribution of the times taken for judges to make comparisons in the Nottinghamshire study. }
    \label{fig: time taken}
\end{figure}

The distribution of the times to make each comparison is shown in Figure \ref{fig: time taken} and the median time taken by each judge is reported in Table \ref{tab: number comparisons}. We see that all judges took a few seconds to make the comparisons, and across all comparisons the median time was four seconds. On this basis, we have no reason to exclude any judge.  Despite being asked to spend ten minutes making comparisons, judge 12 spent almost two hours making comparisons and made more comparisons than all of the other judges combined.  We carried out a sensitivity analysis of their comparisons and describe the results in full in the Supplementary Material. We found their comparisons to be in agreement with the rest of the judges. 

\subsection{Risk of Forced Marriage in Nottinghamshire}
We fitted the BSBT model using the P\'olya-Gamma latent variable representation for 5,000 iterations, removing the first 50 iterations as a burn-in period. This took 54 seconds on a 2019 iMac with a 3 GHz CPU. We fixed the parameters for the inverse-Gamma prior distribution as $\chi = \omega = 0.1$. Trace plots were examined to ensure the Markov chain had converged and mixed well, and are shown in the Supplementary Material. 

\begin{figure}[h]
    \centering
    \includegraphics[width = 0.48\textwidth]{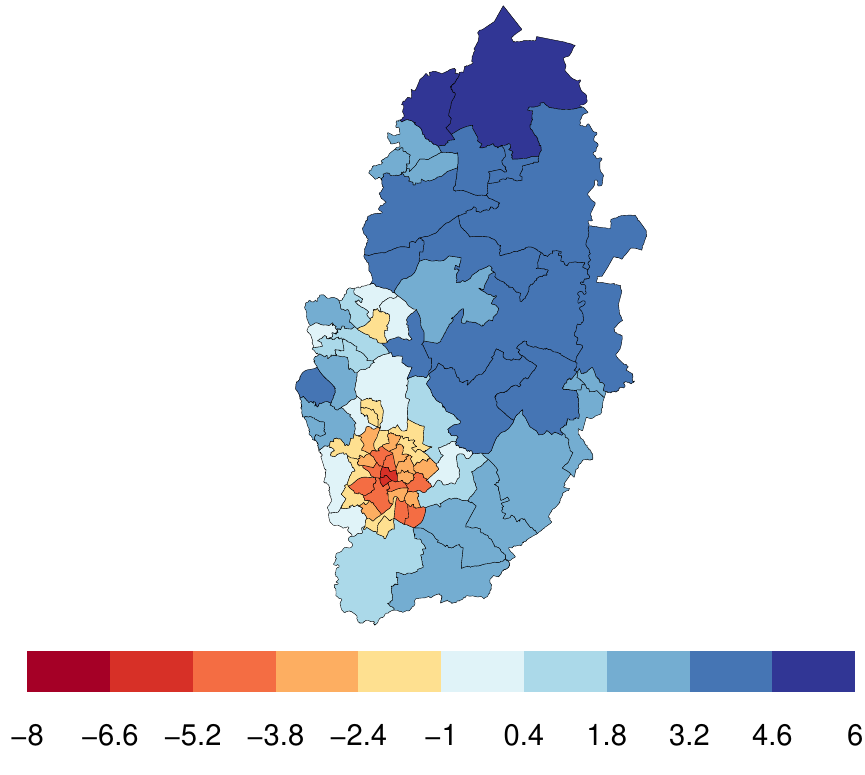}
    \includegraphics[width = 0.48\textwidth]{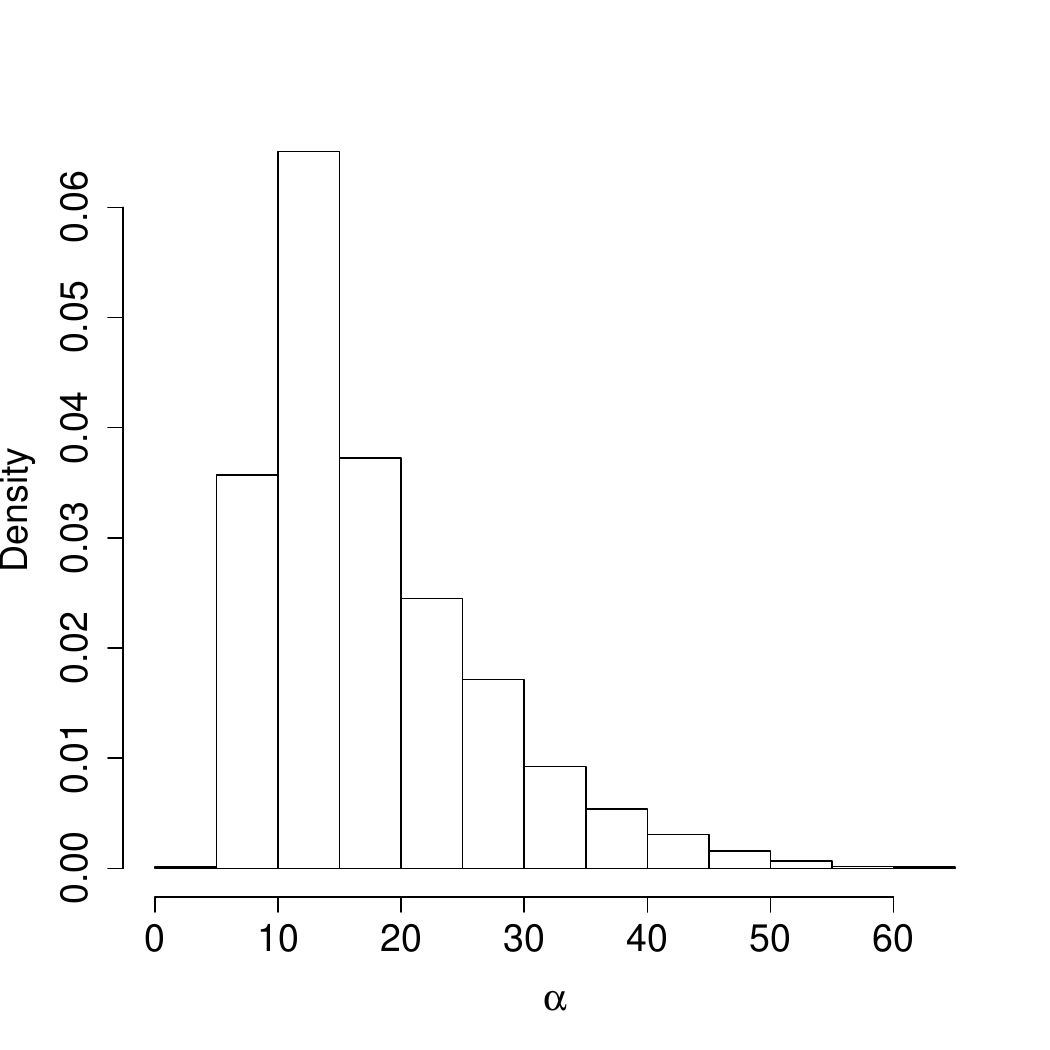}
    \caption{Left: A map of the wards in Nottinghamshire showing the results of the forced marriage study. The wards are coloured by the posterior median estimate for the risk for forced marriage. The wards with the highest risk are coloured in red, the lowest in blue. Right: The posterior distribution for the hyperparameter $\alpha$.}
    \label{fig: Notts map}
\end{figure}

\begin{figure}[h]
    \centering
    \includegraphics[width = 0.48\textwidth]{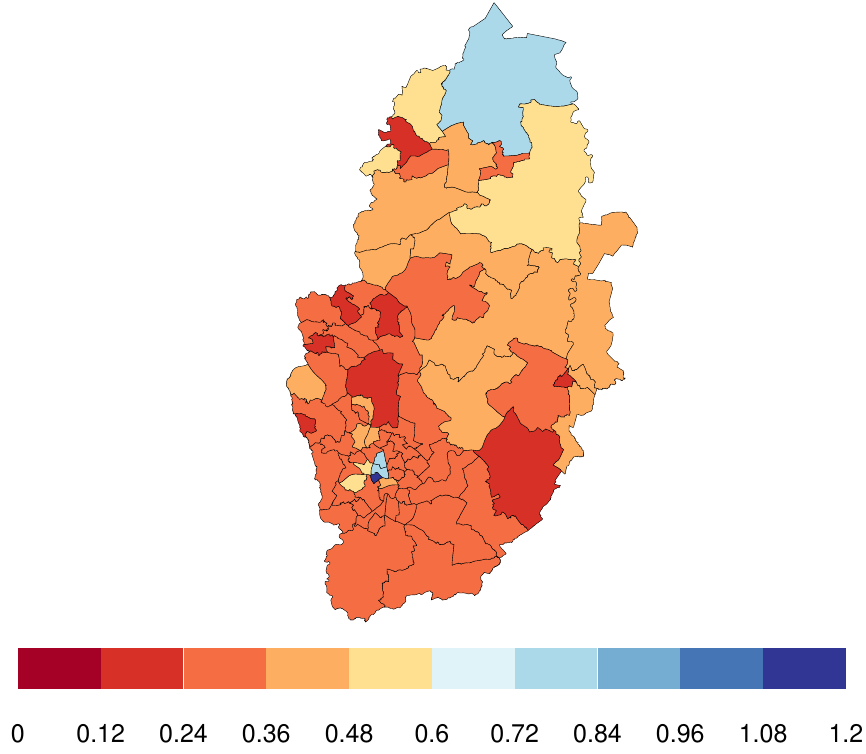}
    \includegraphics[width = 0.48\textwidth]{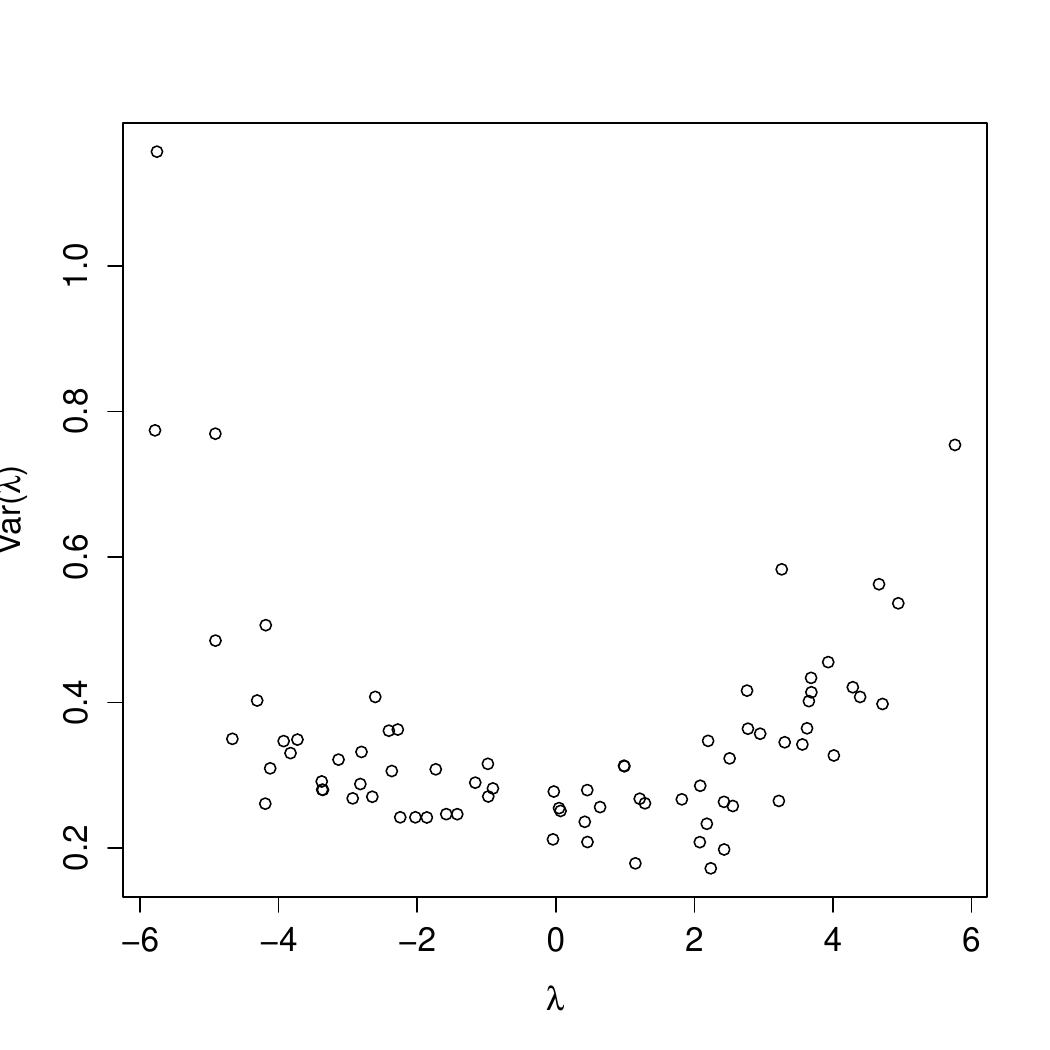}
    \caption{The uncertainty in the results for the Nottinghamshire forced marriage study. Left: The wards are coloured by the variance in the posterior distribution. Right: The posterior median estimate for risk of forced marriage against the variance of the posterior distribution.}
    \label{fig: Notts variance}
\end{figure}

The posterior median estimates for risk of forced marriage are shown in Figure \ref{fig: Notts map}. We see a divide between urban and rural areas in the county, with the ten wards with the highest rates all in the city of Nottingham. Mansfield is the other main urban centre in the county that is noticeable in the results. However, not all urban centres in the county are estimated to have a high risk of forced marriage with Worksop and Newark both ranked in the bottom third of wards.  This suggests the judges were not making judgements based on population size. The posterior distribution for the covariance hyperparameter $\alpha$ is also shown in Figure \ref{fig: Notts variance}; the posterior median is 14.8 with 95\% credible interval (7.39, 40.8). A sensitivity analysis for the covariance hyperparameter can be found in the Supplementary Material. 

Figure \ref{fig: Notts variance} shows the posterior variance for each ward and the posterior variance compared against the posterior median. There is a clear relationship between the posterior median and variance of the risk parameters, where wards with very high or low risk have the largest posterior variance. This suggests some disagreement between these judges on the magnitude of these more extreme risk parameters.

\subsection{Sensitivity Analysis for Judge Twelve}
Judge 12 made over 1,000 comparisons over a period spanning one hour 40 minutes. This is a vastly higher number of comparisons than any other judge, representing around 60\% of the total number of comparisons. We carried out a sensitivity analysis to assess the effect of this judge's comparisons and as a form of quality assurance for our analysis. We followed the split-half method for comparative judgement described in \cite{Bisson2016}. For the analysis, we fitted the model to three data sets: \textit{All} (all the comparisons), \textit{Excluding 12} (all the comparisons excluding judge 12) and \textit{Only 12} (only the comparisons from judge 12).

We fitted the model to the three data sets in the same manner as to the full data set, running the MCMC algorithm for 5,000 iterations. Table \ref{tab: judge 12 corr} shows the Pearson and Spearman's rank correlation coefficients between the estimates for the risk parameters $\boldsymbol{\lambda}$ for each pair of data sets. Both coefficients show a high level of agreement between judge 12 and the other judges. Removing the comparisons from judge 12 has little effect on the ordering of the wards (Spearman's rank correlation coefficient 0.933). 

\begin{table}
    \centering
    \begin{tabular}{|c|c|c|} \hline 
        Data set & All & Excluding 12  \\ \hline
        Excluding 12 & 0.923 & - \\
        Only 12 & 0.971 & 0.832  \\  \hline
    \end{tabular}
    \qquad
\begin{tabular}{|c|c|c|} \hline
Data set & All & Excluding 12  \\ \hline
Excluding  12 & 0.933 & - \\
Only 12 & 0.976 & 0.861  \\ \hline
\end{tabular}
\caption{The correlation coefficient (left) and Spearman's rank correlation coefficient (right) between the estimates for the risk parameters $\lambda$.}
\label{tab: judge 12 corr}
\end{table}

\begin{figure}
    \centering
    \includegraphics[width = \textwidth]{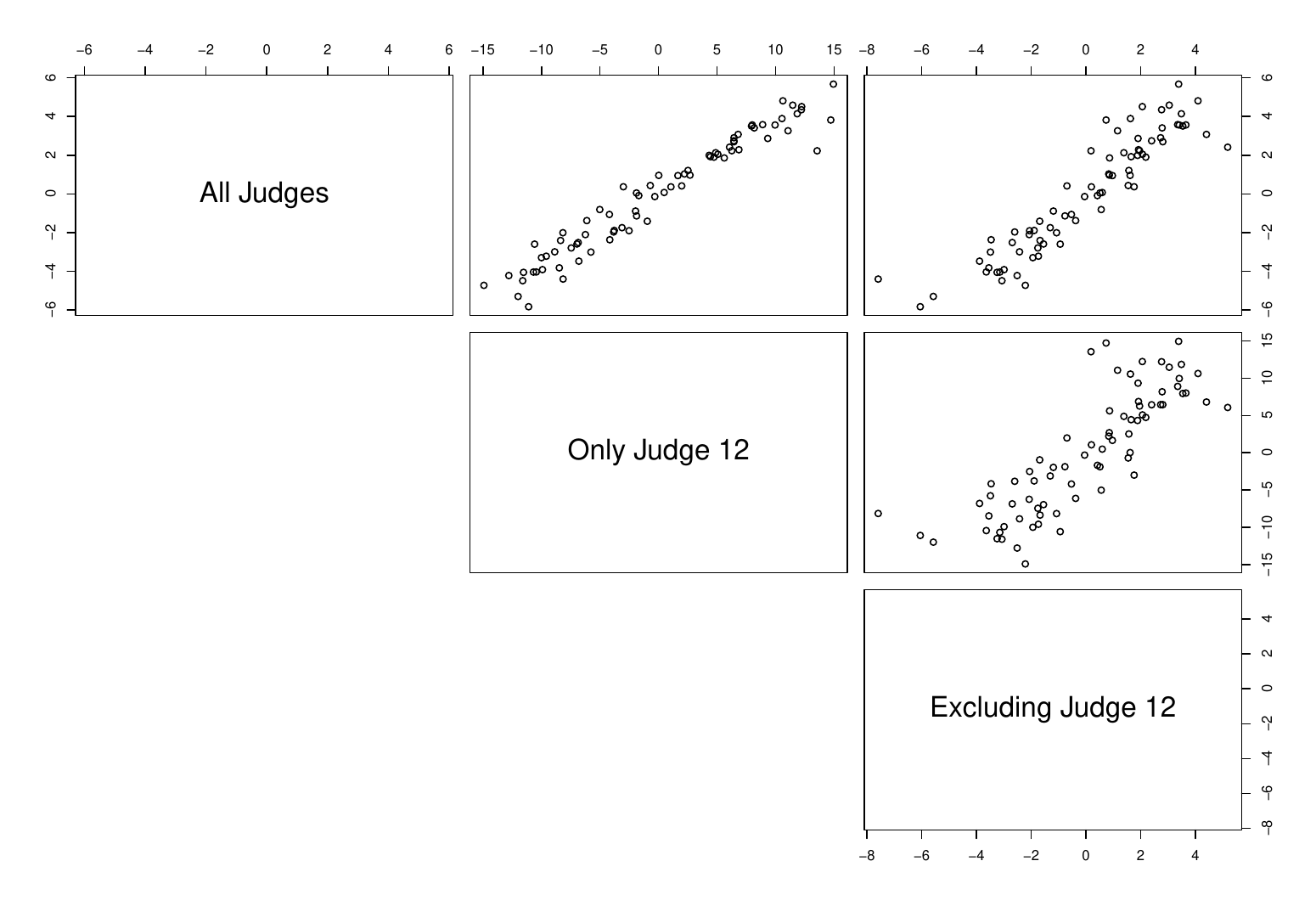}
    \caption{Scatter plots showing the estimated risk of forced marriage using each of the three subsets of the data set. }
    \label{fig:sensitivity scatterplot}
\end{figure}

\begin{figure}
    \centering
    \includegraphics[width = 0.45\textwidth]{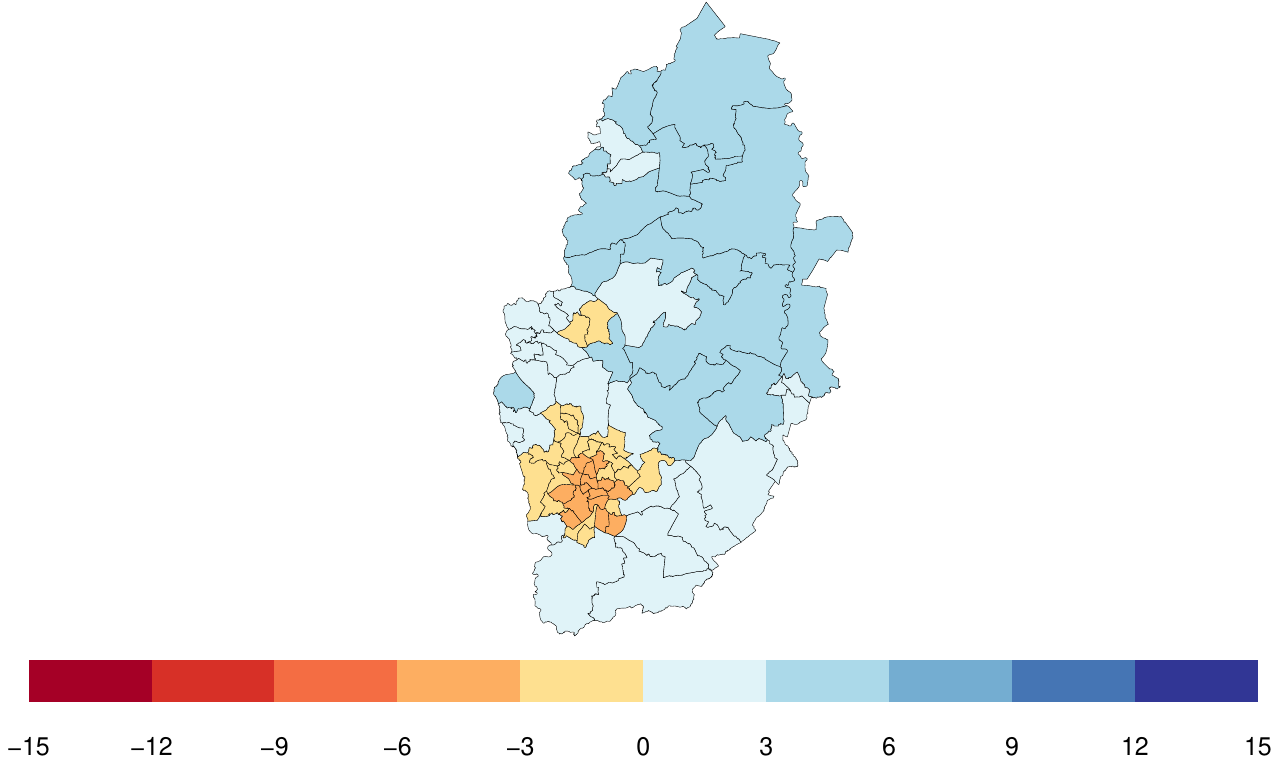}
    \includegraphics[width = 0.45\textwidth]{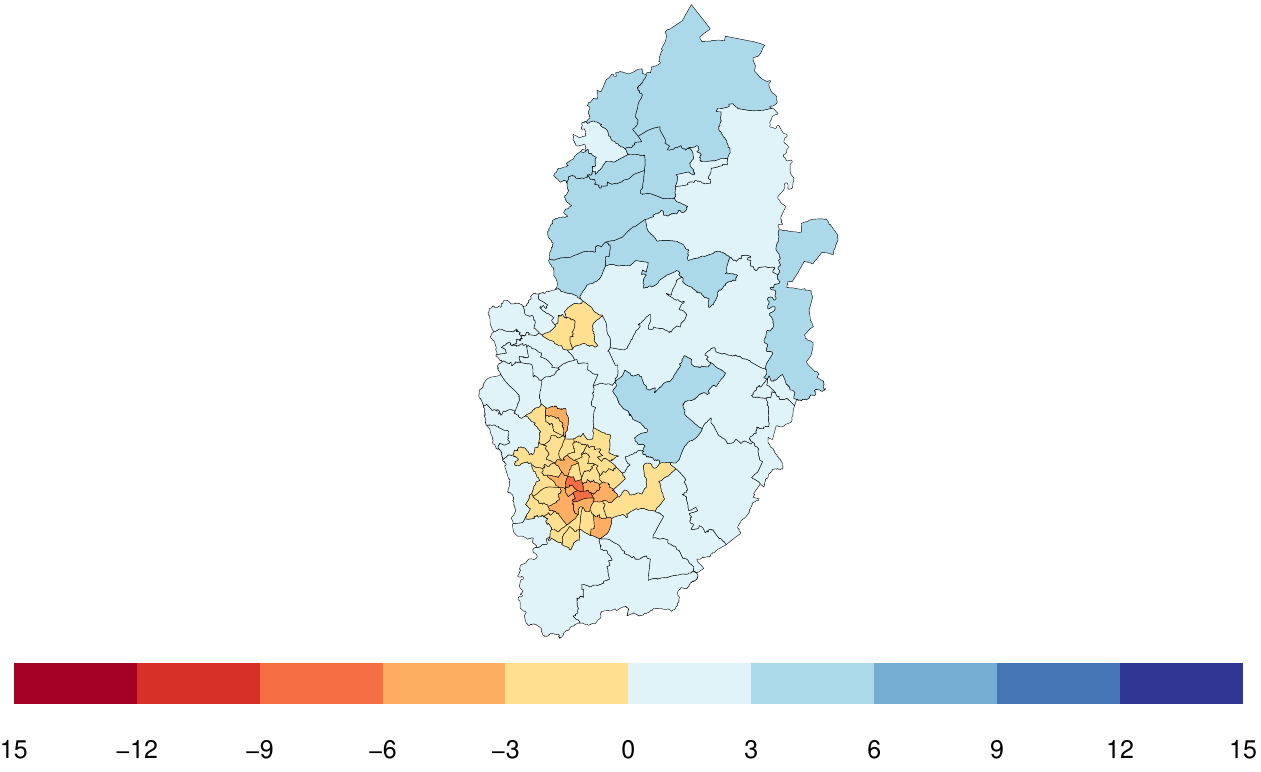}
        \includegraphics[width = 0.45\textwidth]{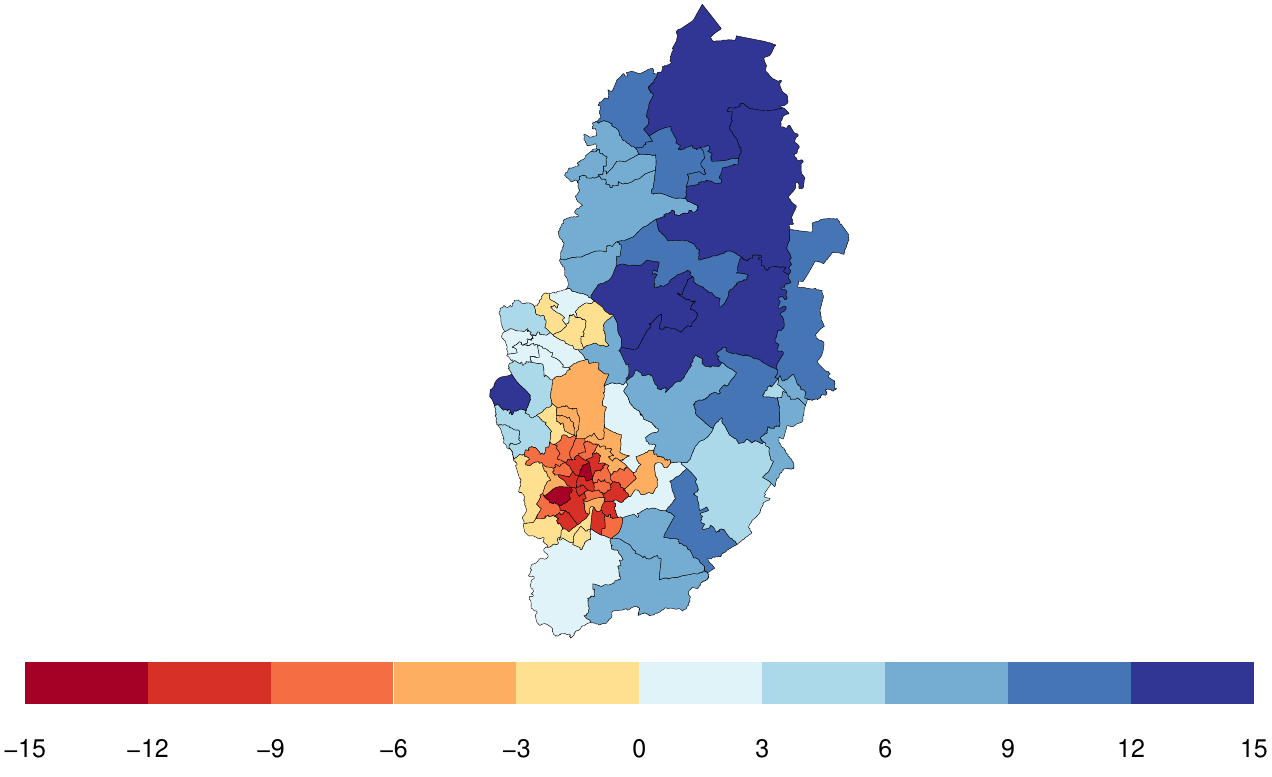}
    \caption{Maps of Nottinghamshire colour coded according to the posterior median estimates from each data set. The map on the top left uses comparisons from all judges, the top right map uses comparisons from all judges excluding judge 12, and the bottom map uses comparisons only from judge 12.}
    \label{fig:sensitivity map}
\end{figure}

Figure \ref{fig:sensitivity scatterplot} shows the posterior median estimates for the risk of forced marriage using each of the three data sets.  Figure \ref{fig:sensitivity map} shows these estimates again, but visualised on maps of the county. The relationship is highly linear showing agreement between the judges on both where high/low risk areas are and the ordering of the wards by prevalence. Of particular importance for our conclusion, is that all judges agree on which areas have the highest risk of forced marriage. The scale of the estimates is different for judge 12, however this does affect the conclusions as we are not able to interpret individuals parameter estimates but their position relative to other estimates. 

\subsection{Clustering for Forced Marriage in Nottinghamshire}
We fitted the spatial clustering BT model described in Section \ref{sec: clustering} to identify clusters of wards that have similar levels of forced marriage and are nearby. We described the spatial relationship between the wards through the  matrix exponential of the county's adjacency matrix, i.e. $f(i, j) = (\exp A)_{ij}$, which matches the prior distribution in the BSBT model. We ran the MCMC algorithm for 100,000 iterations, removing the first 1,000 iterations as burn-in period. We fixed the concentration parameter $\beta = 1\times 10^{-8}$ based on \cite{Ghosh11}.  We ran a sensitivity analysis on the effect of different values of $\beta$, which is described in the Supplementary Material, and found this value had very little impact on the results. Each run of the MCMC algorithm took 45 minutes on a 2019 iMac with a 3 GHz CPU. 

\begin{figure}
    \centering
    \includegraphics[width = 0.2\textwidth]{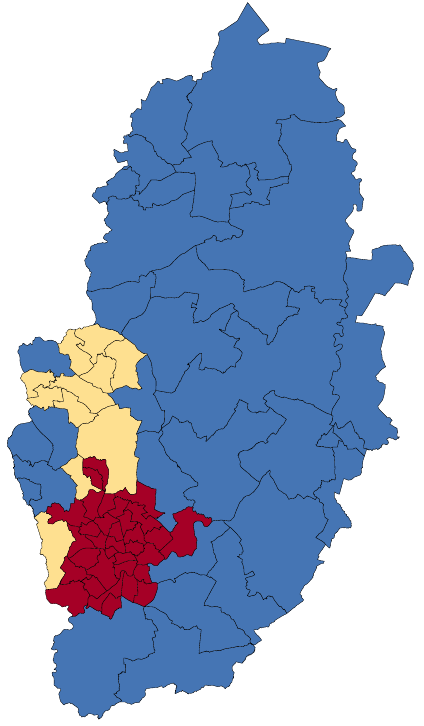}
    \includegraphics[width = 0.32\textwidth]{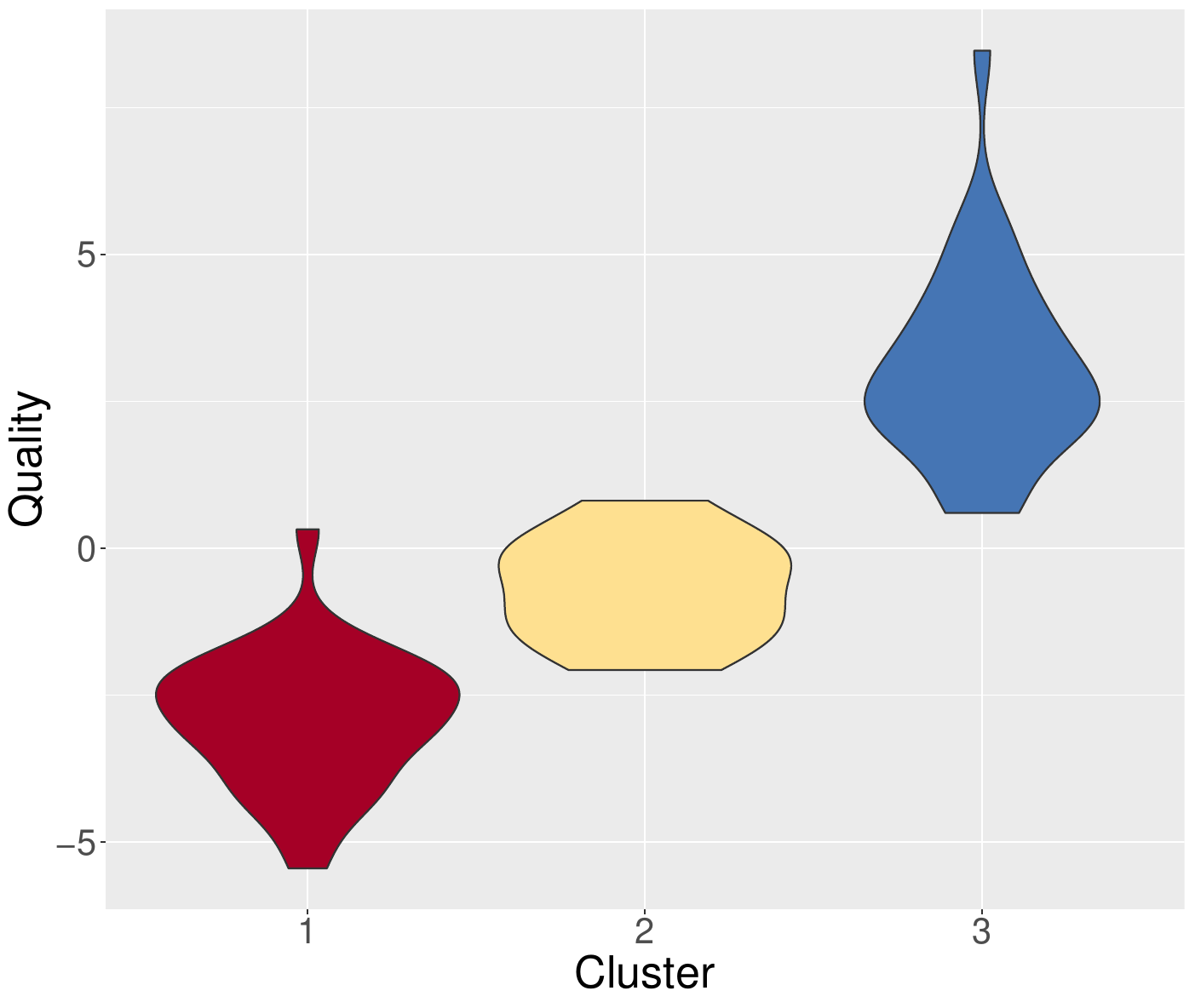}
    \includegraphics[width = 0.32\textwidth]{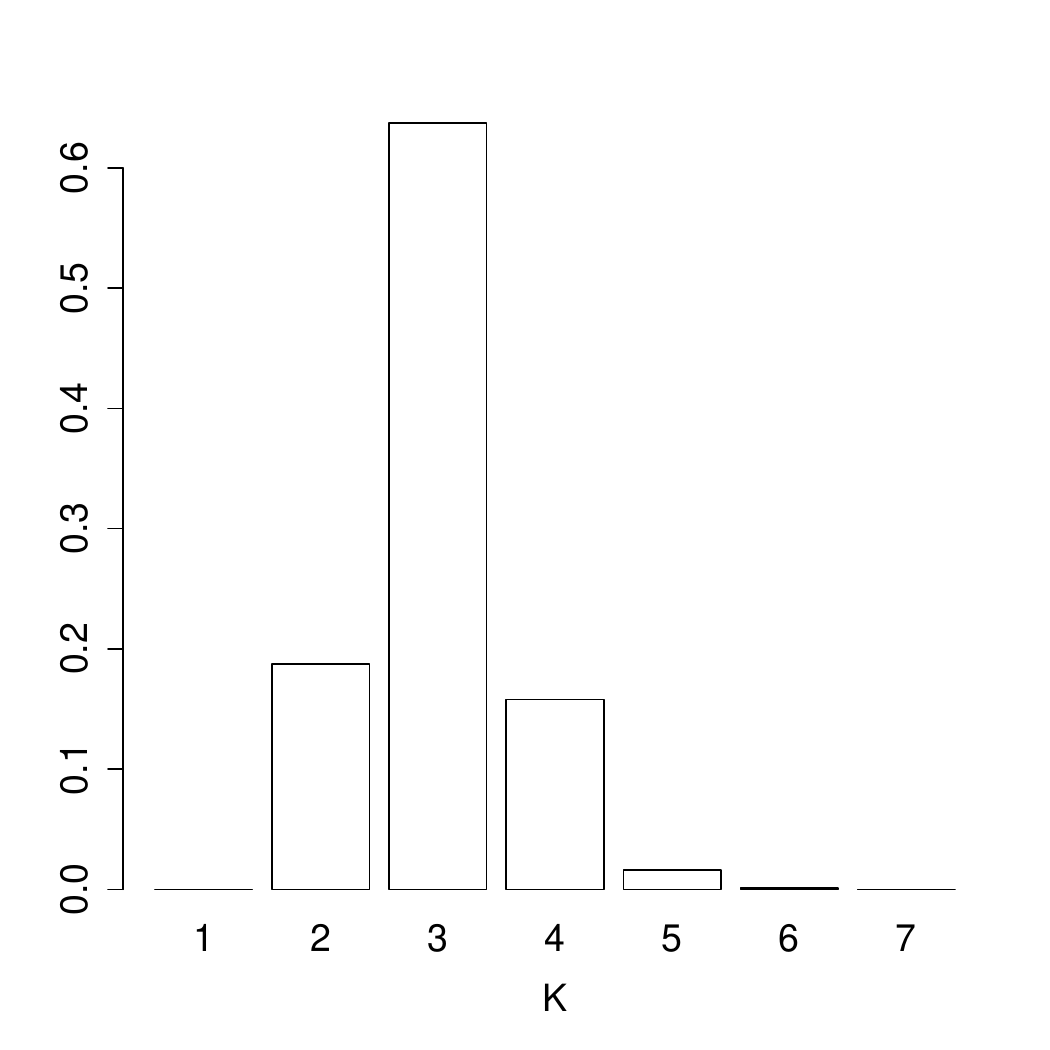}
    \caption{Left: A map of Nottinghamshire with the wards clustered by geolocation and risk of forced marriage. Middle: Violin plots showing the distribution of the risk in each cluster. Right: The posterior probabilities for the number of clusters in the county. }
    \label{fig: Notts clusters}
\end{figure}

The model identified three clusters of wards in the county and these are shown in Figure \ref{fig: Notts clusters}, alongside violin plots for the risk of forced marriage in the clusters and the posterior distribution for the number of clusters. There was little uncertainty about the number of clusters, with the model assigning a probability of 0.637 to there being three clusters, 0.188 to two clusters and 0.156 to four clusters. In the case of two clusters, clusters one and two (red and yellow in Figure \ref{fig: Notts clusters}) were merged, and in the case of four clusters the Nottingham urban area were split into inner and outer city clusters. 

Focusing on the three cluster model, the clusters can be interpreted both as high-medium-low risk areas and through their geolocation. Cluster one consists of wards with high rates of forced marriage and broadly aligns with the Office for National Statistics definition of the Nottingham urban area. The second cluster consists of wards with middling rares of forced marriage and aligns with the Mansfield urban area and a nearby Nottingham suburb. The third cluster is the remainder of the county and forms a cluster of wards with low rates of forced marriage.

\begin{figure}
    \centering
    \includegraphics[width = 0.48\textwidth]{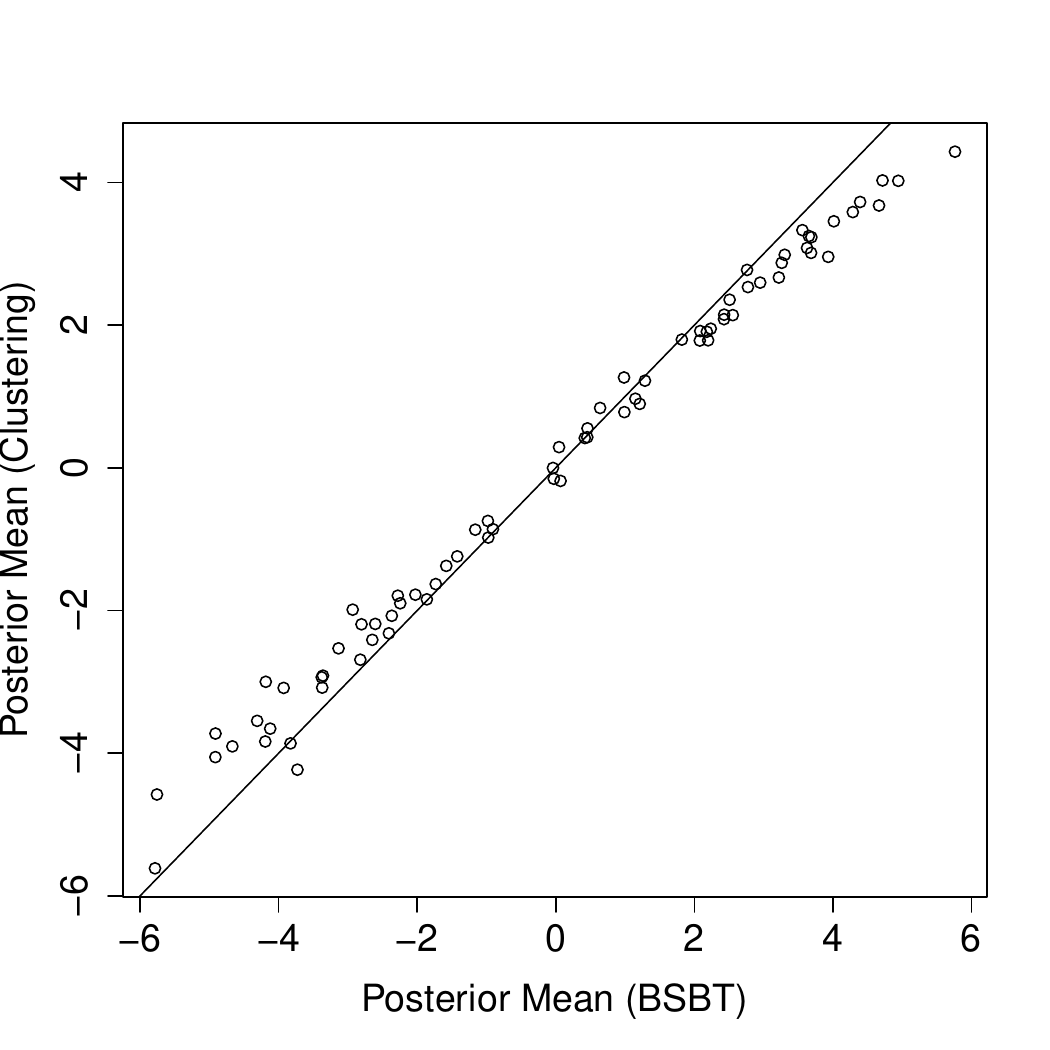}
    \includegraphics[width = 0.48\textwidth]{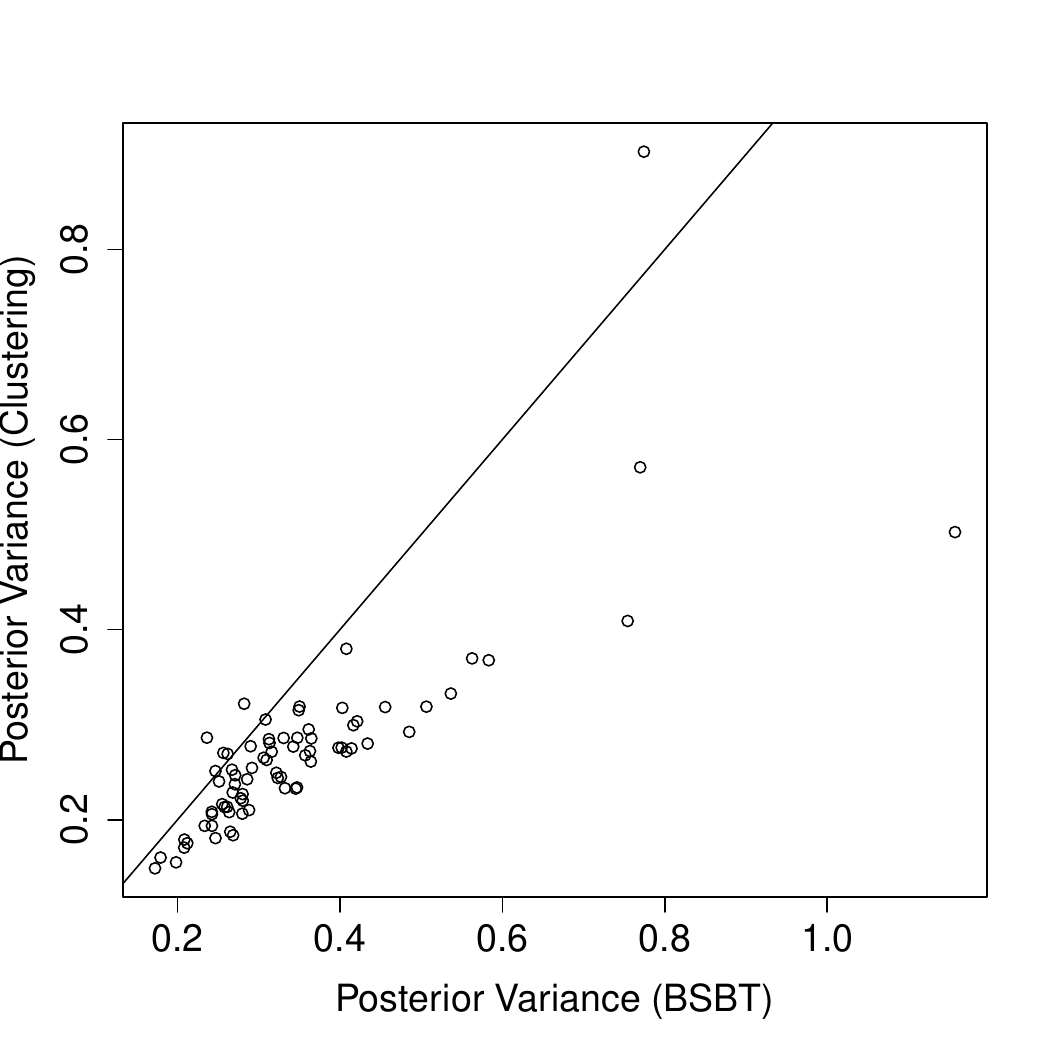}
    \caption{Scatter plots comparing the posterior medians (left) and posterior variances (right) of the BSBT and spatial clustering model.}
    \label{fig: Model comparison}
\end{figure}

Figure \ref{fig: Model comparison} compares the results from the BSBT model to the clustering model. We found little substantive difference between the posterior medians. The clustering model showed some slight shrinkage in the posterior medians compared to the BSBT model. The clustering model however did show reduced uncertainty in the estimates. One reason for this may be because the concentration parameter $\beta$ in the clustering model was fixed, whereas the signal variance parameter $\alpha^2$ in the BSBT model was inferred. Overall, there were marginal differences in the estimates from both models and the clustering model provides results that were easier for our partners in the council to interpret compared to the BSBT model. We also fitted the standard model (i.e. with no spatial correlation) and found that the posterior median estimates for the risk parameters are similar, but the variance much larger. More details can are presented in the Supplementary Material. 

\section{Discussion}
Identifying high risk areas for forced marriage is an important step to prevent people from being forced into marriages. Current data on forced marriage in the UK is not available at a local level, which limits the interventions that safeguarding professionals can design and implement. Comparative judgement studies are well suited to generate risk maps with high spatial resolution, but are limited to applications where are large number of people can be called upon as judges. By developing an efficient scheduling mechanism, we were able to collect comparative judgement data from a limited number of judges. This allowed us to generate ward level risk estimates for forced marriage in the county of Nottinghamshire, UK. Using Bayesian nonparametric methods, we were able to cluster the wards by both risk and geolocation, providing information on forced marriage to support the county's Strategy on Tackling Violence Against Women and Girls. 

In partnership with Nottingham Safeguarding Children's Partnership, which provides the safeguarding arrangements required under the Children and Social Work Act (2017) and the statutory guidance ‘Working Together to Safeguard Children 2018’, we developed a safeguarding training programme for forced marriage in the county. So far, this course has been run three times and attended by 130 safeguarding professionals (110 in Nottinghamshire and 20 in Nottingham), including teachers, health professionals and social workers \citep[p.9, ][]{NSCP24}. We developed a policy briefing (see Supplementary Material) based on our findings and shared this widely with practitioners across the county. We also ran three in-person and online briefings with attendance from councillors, police officers, local authority support staff, housing agency workers and faith leaders. We showed how our findings can be used to inform safeguarding practices for forced marriage. One of the project team appeared as an expert witness in front of the House of Lords Modern Slavery Act 2015 Committee to discuss this work and how it could inform training opportunities and improve data collection about this crime \citep{HoL24}. Our risk maps identify wards in Nottinghamshire where front-line local authority workers should be alerted to the increased risk of forced marriage.  Research has shown that schools and colleges can act as a support network for potential victims \citep{Khan21} and schools and colleges in wards we identified as high risk can work with social services to support potential victims. 

Our methods can also be used to improve forced marriage statistics nationally. We ran two in-person sessions with Members of the UK's Parliament (MPs) and wrote to the Minister for Safeguarding to discuss these findings and improving national forced marriage statistics. MPs were then able to table questions in parliament asking the Government to improve its reporting on forced marriage statistics and making data available at ward level \citep[see, e.g.,][]{NadiaQ, LilianQ}. Our research findings were also accepted as evidence to the House of Commons Committee of Public Accounts' inquiry into protecting vulnerable adolescents \citep{PAC_evidence} and the Women and Equalities Committee's inquiry into so-called honour based abuse \citep{WEC_evidenice}. In October 2022, the Minster for Safeguarding wrote to say ``the research you have shared complements the wider work of the Government in this area" and the Government has ``committed to explore options to better understand the prevalence of forced marriage". We shared our results with the United Nation's Special Rapporteur on extrajudicial, summary or arbitrary executions as part of his call for input on the investigation and documentation of gender-based killings of women and girls. 

There are several limitations to our findings. We lacked sufficient data or information about the judges to carry out any judge specific analysis. Future work could model the effect of each judge, investigating their internal reliability. We also required the judges to either make a decision for each comparisons or skip it. Introducing ties may allow us to learn more information when wards with similar risk levels are compared.  More practically, although we collected over 1,800 comparisons using our online interface, it lacked a counter for the number of comparisons made, which led to one judge providing many more comparisons than others. Future studies should consider either limiting the number of comparisons a single judge can make or showing them a counter with the number of comparisons made so far and a recommended maximum number of comparisons. The minimum number of comparisons required to run a comparative judgement study is still an open question and methodological work on posterior contraction rates may provide clearer estimates for the amount of data that needs to be collected.

We have described two ways in which the risk parameters can be spatially smoothed. There are several other methods by which this can be achieved. One way is to take an approach based on a generalized additive model \citep{Wood2004} where one could assume that the risk in ward $i$ depends on its spatial position, i.e. $\lambda_i = f(\boldsymbol{x}_i)$ where $\boldsymbol{x}_i$ is the spatial location of ward $i$. This functional form can be modelled paramterically or nonparametrically. \citet{BSBT} discuss modelling the spatial smoothing for BT models using both Euclidean and network based metrics. A mixture of functions was used in \citet{Seymour23} with a Gaussian process prior distribution on each function to model the spatial variation at different resolutions. Another way is to encourage spatial smoothing by including a spatially dependent penalty term. The fused lasso method \citep{Tibshirani2004} is one way of achieving this, where risk parameters are estimated by maximising the likelihood function subject to the constraints $\sum_{i=1}^N|\lambda_i|< s_1$ and $\sum_{i=2}^N|\lambda_i - \lambda_{i-1}| < s_2$, where $s_1$ and $s_2$ are suitably chosen values. The second constraint will smooth the complete set of risk parameters, however this may result in the level of risk in wards with extremely (high or low) levels of risk being underestimated. Instead, it may be possible to incorporate local smoothing through the network structure. In \cite{Ohishi19}, the authors propose a locally smoothing second constraint, where $\sum_{i=1}^N \sum_{j \in D_i}|\lambda_i - \lambda_j| < s_2$, where $D_i$ is the set of wards that are adjacent to ward $i$. A systematic comparison between these different approaches is the subject of future work.

Our study showed that comparative judgement can be used as a tool to map human rights abuses, such as forced marriage, at a high spatial resolution. This can provide local level information on human rights abuses and inform the design and implementation of safeguarding interventions. Bayesian computational methods are able to reduce the need for a large number of study participants and provide relevant results to local stakeholders.

\begin{acks}[Acknowledgments]
We thank Emilia Seminerio for her support with data collection.
\end{acks}

\begin{funding}
This work was supported by the Engineering and Physical Sciences Research Council [grant reference EP/R513283/1], the Economic and Social Sciences Research Council [ES/V015370/1], a UKRI Future Leaders Fellowship [MR/X034992/1], and the Research England Policy Support Fund.
\end{funding}

\begin{supplement}
\stitle{Supplementary Material}
\sdescription{A document describing the results of simulation studies and diagnostics for the Nottinghamshire study. }
\end{supplement}
\begin{supplement}
\stitle{Policy briefing}
\sdescription{A policy briefing describing the results of the study.}
\end{supplement}
\begin{supplement}
\stitle{Code and data}
\sdescription{R code and data to replicate the results.}
\end{supplement}


\bibliographystyle{imsart-nameyear} 
\bibliography{bibliography}       

\end{document}